\documentclass[11pt,oneside]{article}
\usepackage{a4wide}
\usepackage{mathrsfs}
\usepackage{latexsym,bm}
\usepackage{graphicx}
\usepackage{indentfirst}
\usepackage{slashed}
\usepackage{amsmath}
\usepackage{amssymb}
\usepackage{color}
\usepackage{hyperref}
\usepackage{epsfig}
\usepackage[titletoc]{appendix}
\usepackage{multirow}%
\usepackage{rotating}
\usepackage{cite}
\usepackage{ulem} 
\usepackage[top=2.9cm,bottom=2.5cm,left=2.8cm,right=3cm]{geometry}
\usepackage{tabularx}

\newcommand{\be}{\begin{equation}} \newcommand{\ee}{\end{equation}}
\newcommand{\ba}{\left(\begin{array}{c}}
\newcommand{\ea}{\end{array}\right)}
\newcommand{\bea}{\begin{eqnarray}} \newcommand{\eea}{\end{eqnarray}}

\newcommand{\al}{&\!\!\!\!}

\newcommand{\bma}{\left(\begin{matrix}}
\newcommand{\ema}{\end{matrix}\right)}
\newcommand{\bqa}{\begin{eqnarray}}
\newcommand{\eqa}{\end{eqnarray}}
\newcommand{\bqaa}{\begin{eqnarray*}}
\newcommand{\eqaa}{\end{eqnarray*}}
\newcommand{\mL}{\mathcal{L}}

\newcommand{\nn}{\nonumber}

\begin{document}
\thispagestyle{empty}
\title{
\Large \bf Chiral study of the {\boldmath$a_0(980)$} resonance and {\boldmath$\pi\eta$} 
scattering phase shifts in light of a recent lattice simulation }
\author{\small Zhi-Hui Guo$^{a,b}$, \,\,  Liuming Liu$^{b}$,  \,\,
Ulf-G. Mei{\ss}ner$^{b,c}$, \,\, J.~A.~Oller$^{d}$, \,\, A.~Rusetsky$^{b}$   \\[0.5em]
{ \small\it ${}^a$  Department of Physics, Hebei Normal University,  Shijiazhuang 050024, China}\\[0.3em]
{\small\it  $^b$Helmholtz-Institut f\"ur Strahlen- und Kernphysik and
Bethe Center for Theoretical Physics,}\\
{\small\it Universit\"at Bonn, D--53115
Bonn, Germany}\\[0.3em]
{\small\it  $^c$Institute for Advanced Simulation, Institut f{\"u}r
Kernphysik and J\"ulich Center for Hadron Physics,}\\
{\small\it Forschungszentrum  J{\"u}lich, D-52425 J{\"u}lich, Germany}  
\\[0.3em]
{\small {\it $^d$Departamento de F\'{\i}sica. Universidad de Murcia. E-30071 Murcia. Spain}}
}
\date{}

%
\maketitle
\begin{abstract} 
We investigate the $a_0(980)$ resonance within chiral effective field theory through a three-coupled-channel 
analysis, namely $\pi\eta$, $K\bar{K}$ and $\pi\eta'$. A global fit to recent lattice finite-volume energy 
levels from  $\pi\eta$ scattering and  relevant experimental data on a $\pi\eta$ event distribution and 
the $\gamma\gamma\to\pi\eta$ cross section is performed. Both the leading and next-to-leading-order analyses 
lead to similar and successful descriptions of the finite-volume energy levels and the experimental data. 
However, these two different analyses yield different $\pi\eta$ scattering phase shifts for the 
physical masses for the $\pi, K, \eta$ and $\eta'$ mesons. The inelasticities, the pole positions in 
the complex energy plane and their residues are calculated both for unphysical and physical meson masses.

\end{abstract}
{\small PACS numbers: 12.39.Fe, 13.75.Lb, 12.38.Gc}


\section{Introduction}

The nonperturbative meson-meson dynamics of  low-energy QCD, especially 
in the scalar channels, is one of the 
most challenging research topics in hadron physics. The complexity of the strong meson-meson interactions 
is manifested in many resonances that appear in various scattering processes~\cite{Olive:2016xmw}. 
Well-known examples are 
the $f_0(500)$ (or $\sigma$) in $\pi\pi$ scattering, the $f_0(980)$ in $\pi\pi$ and $K\bar{K}$ 
coupled channels, the $a_0(980)$ in $\pi\eta$ and $K\bar{K}$ scattering, and the 
$K^*_0(800)$ (or $\kappa$) in the $\pi K$ channel. 
Though it seems plausible that the light isoscalars $f_0(500)$ and $f_0(980)$, the isovector $a_0(980)$ 
and the isospin one-half $K^*_0(800)$ may form a nonet~\cite{Oller:2003vf}, the situation for those 
scalars is much less clear than for the vector nonet $\rho(770)$, $K^*(892)$, 
$\omega(778)$ and $\phi(1020)$. 

A reliable way to obtain further insights into these scalar mesons is based on the low-energy 
effective field theory of QCD, chiral perturbation theory ($\chi$PT), and the unitarity and 
analyticity requirements of the corresponding scattering 
amplitudes~\cite{Dobado:1989qm,Oller:1997ti,Dobado:1996ps,Oller:1998hw,Oller:1998zr}. In this approach, 
one usually needs scattering data, such as the phase shifts or inelasticities, as inputs to 
constrain the free parameters. In the last decade, enormous progress has been made in
the study of the $f_0(500)$, see Ref.~\cite{Pelaez:2015qba} for a recent review. 
It is most likely that the effects from the inelastic channels, such as $K\bar{K}$ and other higher 
ones, are small for the $f_0(500)$. As a result, one can use the single-channel formalism to describe 
this broad resonance well. In addition, many existing precise $\pi\pi$ scattering data also help to 
precisely determine the $f_0(500)$ pole position. The experimental $\pi K$ phase shifts also confirm 
the existence of the $K^*_0(800)$ as a pole in the complex energy 
plane~\cite{DescotesGenon:2006uk,Zheng:2003rw,Pelaez:2016tgi,Dobado:1996ps,Oller:1998hw,Oller:1998zr}. 
Due to the proximity of the $f_0(980)$ and $a_0(980)$ to the $K\bar{K}$ threshold, the coupled-channel 
formalism is essential to study these two states. Rigorous dispersive studies have been performed for 
the $f_0(980)$, see Ref.~\cite{GarciaMartin:2011jx} and references therein. Various unitarized $\chi$PT 
approaches also confirm that there is a well-established resonance pole for the $f_0(980)$ after 
successfully reproducing the $\pi\pi$ scattering data around 
1~GeV~\cite{Oller:1997ti,Oller:1998hw,Oller:1998zr,Albaladejo:2008qa}. However, the situation for 
the $a_0(980)$ is less clear and its pole positions are still under debate~\cite{Oller:1998hw,Oller:1998zr,Albaladejo:2015aca,Wolkanowski:2015lsa,Guo:2011pa,Dudek:2016cru}. 
One of the biggest difficulties in preventing a precise 
determination of the $a_0(980)$ is the lack of direct experimental $\pi\eta$ scattering data. It is 
unlikely that this will be improved in the near future.

Fortunately,  important progress using  lattice QCD simulations for $\pi\eta$ scattering, together with $K\bar{K}$ and $\pi\eta'$ coupled channels, has been made 
very recently~\cite{Dudek:2016cru}. However, the pion mass ($m_\pi \sim 391$~MeV) used in the calculation 
is still much heavier than its physical value. A large number of energy levels in the finite volume 
is obtained 
by using a large amount of interpolating operators and many moving frames. 
These energy levels are then used to extract the $\pi\eta$ phase shifts and inelasticities by 
using L\"uscher's method~\cite{luescher-torus} and parametrizing
the $K$ matrix in various ways\footnote{Note that, recently, the L\"uscher method 
has become a commonly accepted tool to analyze the lattice data in the scattering sector, 
including the case of the multi-channel scattering~(see, e.g., Refs.~\cite{Lage,Wilson,Dudek:2016cru,Prelovsek}).
Different algebraic parameterizations for the $K$-matrix are used and 
the free parameters are fitted to the lattice data on the energy levels. Note also that an alternative approach to study of the inelastic scattering has been formulated recently~\cite{Agadjanov:2016mao}.}. The resulting $\pi\eta$ phase shifts~\cite{Dudek:2016cru} around the $K\bar{K}$ threshold do not show any sharp increase, 
and hence they do not correspond to the behavior of a canonical resonance pole in the complex 
energy plane. Instead the authors of Ref.~\cite{Dudek:2016cru} find that the $a_0(980)$ state corresponds 
to a pole in the fourth Riemann sheet (RS), which is not directly connected to the physical sheet\footnote{The pole on the fourth RS in Ref.~\cite{Dudek:2016cru} lies above the $K\bar{K}$ threshold. The physical sheet in that energy region is directly connected to the third RS, in which a canonical resonance pole should be located. }. 
Although this observation is made with $m_\pi=391$~MeV, interestingly it agrees with the previous 
study in Ref.~\cite{Guo:2011pa} for the $a_0(980)$, which also corresponds to  a fourth RS 
pole. But the calculation in Ref.~\cite{Guo:2011pa} is done with the physical masses for 
the $\pi$, $K$, $\eta$ and $\eta'$ channels.

In order to make a close comparison with the physical $a_0(980)$ state, a proper way to perform 
the chiral extrapolation of the lattice simulations in Ref~\cite{Dudek:2016cru} is essential. 
In this respect, the $\chi$PT framework provides  a reliable tool. In this work we use 
the unitarized $\chi$PT approach  \cite{Oller:1997ti,Oller:1998zr,Oller:2000fj}
to reanalyze the lattice simulations, and then extrapolate the $\pi$, $K$, $\eta$ and $\eta'$ masses 
to their physical values. 
It is worth emphasizing that the methodology for coupled-channel unitarized $\chi$PT in a finite volume 
for the scalar meson sector was developed in 
Refs.~\cite{Bernard:2010fp,Doring:2011vk,Doring:2012eu,Doring:2011nd}. 
Note also that, recently, a similar method was used in Ref.~\cite{Hu:2016shf} in order to extract the position 
of the $\rho$-meson pole from the lattice phase shifts. In particular, it has been argued that the coupling 
to the $K\bar K$ channel might have a significant impact on it.  
Instead of analyzing the phase shifts provided in Ref~\cite{Dudek:2016cru}, we directly fit the 
lattice energy levels by considering  unitarized $\chi$PT in a finite box. 
In addition to the lattice finite-volume energy levels, we also include two kinds of experimental data 
in the global fits, namely, a $\pi\eta$ event distribution~\cite{Armstrong:1991rg} and 
the $\gamma\gamma\to\pi\eta$ cross section~\cite{Uehara:2009cf}, so as to better constrain 
the free parameters in the analyses.   After the successful reproduction of the lattice energy levels 
and experimental data, we then calculate the $\pi\eta$ phase shifts, inelasticities, pole positions, and 
their residues by taking both heavy unphysical and physical masses for $\pi, K, \eta$ and $\eta'$. 

The article is organized as follows. We introduce the unitarized $\chi$PT approach and 
the finite-volume effects in moving frames in 
Sec.~\ref{sec.xptandfv}. The fits to the lattice energy levels and experimental data are analyzed 
in detail in Sec.~\ref{sec.fit}. The $\pi\eta$ scattering phase shifts, inelasticities, 
the $a_0(980)$ and $a_0(1450)$ pole positions and their residues for 
the unphysical masses are given in Sec.~\ref{sec.latdiscuss}. The results after extrapolating 
the $\pi$, $K$, $\eta$ and $\eta'$ masses to their physical values are discussed in 
Sec.~\ref{sec.phydiscuss}. A short summary and conclusions are given in Sec.~\ref{sec.summary}.

{\boldmath
\section{ Unitarized $U(3)$ $\chi$PT and its finite-volume effects}\label{sec.xptandfv}
}

In this section we briefly review the basic aspects of the formalism used to analyze 
lattice QCD energy levels and experimental data.  
Note that $\chi$PT is the effective field theory of low-energy QCD and it has been proven to be quite 
successful to describe the  dynamics of the pseudo-Nambu-Goldstone bosons (pNGBs), including the 
$\pi$, $K$ and $\eta$ mesons~\cite{Gasser:1984gg}. In the present work we study the $a_0(980)$ 
by including the scattering of three coupled channels, namely $\pi\eta$, $K\bar{K}$ and $\pi\eta'$. 
In this case, $U(3)$  $\chi$PT~\cite{HerreraSiklody:1996pm,Kaiser:2000gs} is the proper framework, 
instead of the conventional $SU(3)$ $\chi$PT~\cite{Gasser:1984gg}. This is because the singlet 
$\eta_0$ and the QCD $U(1)_A$ anomaly effect are explicitly included in $U(3)$ $\chi$PT, while in the 
$SU(3)$ case the heavy singlet $\eta_0$ is integrated out. 
The leading-order (LO) Lagrangian of $U(3)$ $\chi$PT reads~\cite{oriua}
\begin{eqnarray} \label{eq.lolagrangian}
\mL_{2}=\frac{ F^2}{4}\langle u_\mu u^\mu \rangle+
\frac{F^2}{4}\langle \chi_+ \rangle
+ \frac{F^2}{3}M_0^2 \ln^2{\det u}\,,
\end{eqnarray}
where $\langle \ldots \rangle$ denotes the trace in flavor space and 
the last term encodes the $U_A(1)$ anomaly effect that gives the singlet $\eta_0$ a 
large mass $M_0$ even in the chiral limit.  The basic chiral operators are defined as   
\begin{eqnarray}
u_\mu &=& i u^+ D_\mu U u^+  \,, \nn\\
\chi_+ &=& u^+ \chi u^+ + u \chi^+ u \,,\nn\\
U &=&  u^2 = \exp\left(i \sqrt2\Phi /F \right) \,, \nn\\
D_\mu U &=& \partial_\mu U - i r_\mu U + i U l_\mu \,, \nn\\
\chi &=& 2 B (s + i p) \,,
\end{eqnarray}
where $F$ denotes  the weak decay constant of the pNGBs in the chiral limit, the parameter 
$B$ is related to the quark condensate through 
$\langle 0|\bar{q}^iq^j|0\rangle =- F^2 B\delta^{ij}$ at leading order, $r_\mu\,, l_\mu \,, s\,, p$ 
are external sources and the pNGBs are collected in the $3\times 3$ matrix  
\begin{equation}\label{eq.phi1} 
\Phi \,=\, \left( \begin{array}{ccc}
\frac{1}{\sqrt{2}} \pi^0+\frac{1}{\sqrt{6}}\eta_8+\frac{1}{\sqrt{3}} \eta_0 & \pi^+ & K^+ \\ \pi^- &
\frac{-1}{\sqrt{2}} \pi^0+\frac{1}{\sqrt{6}}\eta_8+\frac{1}{\sqrt{3}} \eta_0   & K^0 \\  K^- & \bar{K}^0 &
\frac{-2}{\sqrt{6}}\eta_8+\frac{1}{\sqrt{3}} \eta_0 
\end{array} \right)\,.
\end{equation} 
The explicit chiral symmetry breaking is realized by taking the vacuum expectation 
values of the scalar source $s = {\rm diag}(m_u,m_d,m_s)$, with $m_q$ the light-quark masses. 
We work in the isospin symmetry limit $m_u = m_d$. 

The physical $\eta$ and $\eta'$ states result from the mixing of the octet $\eta_8$ and the 
singlet $\eta_0$. At leading order, it is enough to introduce one mixing angle $\theta$ 
to diagonalize the quadratic terms of $\eta_0$ and $\eta_8$ 
\begin{eqnarray}\label{eq.deflomixing} 
\eta_8 &=&  c_\theta \overline{\eta}+ s_\theta\overline{\eta}'\,,\nonumber \\
\eta_0 &=& -s_\theta\overline{\eta} + c_\theta\overline{\eta}'\,, 
\end{eqnarray}
with $c_\theta=\cos\theta$ and $s_\theta=\sin\theta$. Here, we use the notation $\overline{\eta}$ and  
$\overline{\eta}'$ to denote the diagonalized fields of the Lagrangian Eq.~\eqref{eq.lolagrangian} 
at leading order. When higher order contributions are included, $\overline{\eta}$ and $\overline{\eta}'$ 
will get mixed again and we refer to Refs.~\cite{Guo:2011pa,Guo:2015xva,Jamin:2000wn} for further details on 
handling the higher order mixing effects.  
The LO mixing angle $\theta$ can be calculated in terms of the singlet $\eta_0$ mass $M_0$ in the 
chiral limit and the LO masses of the pion and the kaon~\cite{Guo:2011pa,Guo:2015xva}  
\begin{eqnarray}\label{eq.loangle}
\sin{\theta} &=& -\left( \sqrt{1 +
\frac{ \big(3M_0^2 - 2\Delta^2 +\sqrt{9M_0^4-12 M_0^2 \Delta^2 +36 \Delta^4 } \big)^2}{32 \Delta^4} } ~\right )^{-1}\,,
\end{eqnarray}
where $\Delta^2 = \overline{m}_K^2 - \overline{m}_\pi^2$, and $\overline{m}_K$ and $\overline{m}_\pi$ are
the LO  kaon and pion masses, respectively. 

The higher order contributions in $\chi$PT include both the chiral loops and the higher order  
low-energy constants (LECs). In Ref.~\cite{Guo:2011pa} the one-loop calculation of all two-body 
light-meson scattering amplitudes is carried out within  $U(3)$ $\chi$PT. A systematical study of the 
$O(p^4)$ Lagrangian of $U(3)$ $\chi$PT is given in Refs.~\cite{HerreraSiklody:1996pm,Kaiser:2000gs}. 
Another way to account for the effects from the higher order LECs is to include  resonance exchanges in 
a chiral invariant way~\cite{Ecker:1988te}. The pioneering study of resonance exchanges in the chiral 
framework for the $\pi\pi$ and $\pi K$ scattering was given in Ref.~\cite{Bernard:1991zc}. A generalization 
to  include the leading resonance exchanges in the chiral counting in all the meson-meson scattering 
channels is completed in Ref.~\cite{Guo:2011pa}. 
For a more detailed account the reader is referred to Ref.~\cite{Guo:2011pa} and references therein.

\subsection{Brief reminder of unitarized {\boldmath$U(3)$ $\chi$}PT}\label{sec.introu3xpt}

Since $\chi$PT is organized in a double expansion in momenta and light-quark masses, it can 
only be applied for low-energy  processes involving  the pNGBs. In the higher energy region, especially
when the resonances appear, the perturbative $\chi$PT amplitudes start to severely violate the unitarity 
condition and one can not trust the $\chi$PT expressions anymore. The unitarization procedure, which 
restores the unitarity of the perturbative $\chi$PT amplitudes, provides a useful tool to extend 
the $\chi$PT domain to the resonance energy  region. 
However, this is usually done at the expense of violating crossing symmetry, and such a 
unitarization procedure unavoidably introduces some model dependence from the chosen set 
of higher order effects that are resummed. 
In the single $\pi\pi$ channel case, unitarity and analyticity can be strictly implemented within a 
range of energies and  different groups obtain  quite compatible results for the $f_0(500)$ pole 
positions~\cite{Pelaez:2015qba}. However, a rigorous solution for the coupled-channel scattering is 
typically not possible 
and usually different types of approximations are introduced. A convenient way to proceed is to treat 
the right-hand cut (or the  unitarity cut) nonperturbatively, 
whereas the cross-channel effects are included 
in a perturbative fashion~\cite{Oller:1997ti,Oller:2000fj,Oller:1999me}. Indeed, this is the case in many 
unitarized $\chi$PT studies~\cite{Oller:1997ti,Oller:1998hw,Oller:1998zr}.

A unitarization of the perturbative meson-meson scattering amplitudes up to the next-to-leading 
order (NLO) calculated in the  one-loop $U(3)$ $\chi$PT plus tree-level resonance 
exchanges~\cite{Guo:2011pa} 
is then performed using the formalism of Ref.~\cite{Oller:1998zr}.
The final expression for the meson-meson scattering amplitude $\mathcal{T}^I_J(s)$ reads 
\begin{eqnarray}\label{eq.ndt}
\mathcal{T}^I_J(s) =  [1 +  N^I_J(s)\cdot G(s) ]^{-1} \cdot N^I_J(s)  \,, 
\end{eqnarray}
where $I$ and $J$ denote the isospin and angular momentum, respectively.
This unitarization method corresponds to an algebraic approximation of the conventional $N/D$ method 
\cite{Oller:1998zr}.  
By construction, the function  $G(s)$  in Eq.~\eqref{eq.ndt} incorporates the two-body 
right-hand cut and it is given by the standard two-point one-loop function
\begin{eqnarray}\label{eq.defg}
G(s)=i\int\frac{{\rm d}^4q}{(2\pi)^4}
\frac{1}{(q^2-m_1^2+i\epsilon)[(P-q)^2-m_2^2+i\epsilon ]}\ ,\qquad
s\equiv P^2\ \,,
\end{eqnarray}
which can be calculated by  a once-subtracted dispersion relation or in the dimensional regularization 
by replacing the divergence with a constant. The explicit expression of $G(s)$ reads~\cite{Oller:1998zr}
\begin{eqnarray}
\label{eq.gfuncdr}
G(s)^{{\rm DR}} \al=\al\frac{1}{16\pi^2}\bigg\{{a}(\mu)+\ln\frac{m_1^2}{\mu^2}
+\frac{s-m_1^2+m_2^2}{2s}\ln\frac{m_2^2}{m_1^2}\nonumber\\
\al\al+\frac{\sigma(s)}{2s}\bigg[\ln\big(\sigma(s)+s-m_2^2+m_1^2\big)-\ln\big(\sigma(s)-s+m_2^2-m_1^2\big)\nonumber\\
\al\al+\ln\big(\sigma(s)+s+m_2^2-m_1^2\big)-\ln\big(\sigma(s)-s-m_2^2+m_1^2\big)\bigg]\bigg\}\ ,
\end{eqnarray}
where the superscript ${\rm DR}$ denotes the use of the dimensional regularization in the
expression of $G(s)$,  
$m_1$ and $m_2$ are the masses of the two intermediate mesons in the scattering process, 
$a(\mu)$ is the subtraction constant and 
\begin{eqnarray}\label{eq.defsigkin}
\sigma(s)=\sqrt{\lambda(s,m_1^2,m_2^2)}\,,
\end{eqnarray} 
with $\lambda(a,b,c)=a^2+b^2+c^2-2ab-2bc-2ac$  the K\"all\'{e}n function. 
The function $G(s)$ itself is independent of the renormalization scale $\mu$, since the explicit 
$\mu$ dependence in the second term in Eq.~\eqref{eq.gfuncdr} is compensated by the 
subtraction constant $a(\mu)$. We will fix $\mu=770$~MeV in our later discussion, 
indicating that the value of the subtraction constant obtained here refers to that scale. 
The subtraction constants could be reabsorbed gradually when higher orders are included in 
the function $N(s)$. As it is clear from Eq.~\eqref{eq.ndt}, for an exact elimination of the 
subtraction constants, the inverse of the new function 
$\widetilde{N}(s)$ would be $\widetilde{N}_J^I(s)^{-1}=N_J^I(s)^{-1}+ \delta G$, with $\delta G$ 
the diagonal matrix including only the subtraction constants.

In contrast, the  $N^I_J(s)$ function is free of any two-body 
right-hand cut singularity\footnote{Except those from the channels with heavier thresholds.} 
and only contains the crossed-channel cuts. Its explicit expression is given by~\cite{Guo:2011pa} 
\begin{eqnarray}\label{eq.defnfunction}
N_J^I(s) = {T_J^I}(s)^{\rm (2)+Res+Loop}+ T_J^I(s)^{(2)} \cdot   G(s)  \cdot T_J^I(s)^{(2)} \,, 
\end{eqnarray}
where $T_J^I(s)^{\rm (2)+ Res + Loop}$ are the partial-wave projected $U(3)$ $\chi$PT amplitudes, and the 
superscripts  ``${\rm(2),~ Res}$'' and ``Loop'' denote the LO amplitudes, resonance exchanges and loop 
contributions, respectively. The explicit calculations of these perturbative amplitudes are 
given in detail in Ref.~\cite{Guo:2011pa} and we briefly recapitulate the main results here. 
The LO $S$-wave amplitudes in the isospin $I=1$ channel are 
\begin{eqnarray}\label{eq.pwlo}
 T_{J=0}^{I=1, \pi\eta\to\pi\eta}(s)^{(2)} &=& \frac{(c_\theta-\sqrt2 s_\theta)^2m_\pi^2}{3 F_\pi^2}\,,\nonumber \\
 T_{J=0}^{I=1, \pi\eta\to K\bar{K}}(s)^{(2)} &=& \frac{c_\theta(3m_\eta^2+8 m_K^2 + m_\pi^2-9s)+2\sqrt{2}s_\theta(2m_K^2+m_\pi^2) }{6\sqrt{6} F_\pi^2}\,,\nonumber \\
  T_{J=0}^{I=1, \pi\eta\to\pi\eta'}(s)^{(2)} &=& \frac{(\sqrt{2}c_\theta^2-c_\theta s_\theta - \sqrt2 s_\theta^2)m_\pi^2}{3 F_\pi^2}\,,\nonumber \\
 T_{J=0}^{I=1, K\bar{K}\to K\bar{K}}(s)^{(2)} &=& \frac{s}{4 F_\pi^2}\,,\nonumber \\
 T_{J=0}^{I=1, K\bar{K}\to \pi\eta'}(s)^{(2)} &=& \frac{s_\theta(3m_{\eta'}^2+8 m_K^2 + m_\pi^2-9s) - 2\sqrt{2}c_\theta(2m_K^2+m_\pi^2) }{6\sqrt{6} F_\pi^2}\,,\nonumber \\
 T_{J=0}^{I=1, \pi\eta'\to\pi\eta'}(s)^{(2)} &=& \frac{(\sqrt2 c_\theta + s_\theta)^2m_\pi^2}{3 F_\pi^2}\,,
\end{eqnarray}
where $c_\theta$ and $s_\theta$ are defined in Eq.~\eqref{eq.deflomixing}.

Concerning the resonance exchanges, we mention that in Ref.~\cite{Guo:2011pa} one multiplet of bare 
octet scalar resonances is included at the Lagrangian level, which is mostly responsible for 
the excited physical scalar states of $f_0(1370), K^*_0(1430)$ and $a_0(1450)$. The bare singlet 
scalar introduced at the Lagrangian level is found to be important for the $f_0(980)$. The other 
scalar resonances, such as $\sigma, \kappa$ and $a_0(980)$, are mainly generated from the 
nonperturbative meson-meson contact interactions. Concerning other higher order effects, such 
as the vector resonance exchanges and light pseudoscalar loop contributions, we refer 
to Ref.~\cite{Guo:2011pa} for further details. 
The unknown parameters in our model, including the resonance couplings and the subtraction 
constants, were determined in  Ref.~\cite{Guo:2011pa} by fitting a large amount of experimental data, 
consisting of the $\pi\pi \to \pi\pi, K\bar{K}$ scattering phase shifts and inelasticities in the 
$IJ=00$ channel~\cite{exp:pipi00}, the $\pi\pi\to\pi\pi$ phase shifts with $IJ=1\,1$ ~\cite{exp:pipi11} 
and $IJ=2\,0$~\cite{exp:pipi20}, the $\pi K\to \pi K$ phase shifts with $IJ=\frac{1}{2}\,0, \frac{3}{2}\,0$ and  $\frac{1}{2}\,1$~\cite{exp:piK} and a $\pi\eta$ event distribution in the 
$IJ=1\,0$ case~\cite{Armstrong:1991rg}.  
Note that there is no available direct experimental data for $\pi\eta$ scattering, but the 
$IJ=10$ partial-wave amplitudes can be tested because of their impact through final-state 
interactions. The $\pi\eta$ event distribution taken corresponds to the measured one 
in the complicated reaction $p p \to p p \eta \pi^+ \pi^-$ ~\cite{Armstrong:1991rg}. 
Since it is quite possible that the $K\bar{K}$ and $\pi\eta$ intermediate states may enter in 
different ways  in $p p \to p p \eta \pi^+ \pi^-$, we introduce two parameters $c_1$ and $c_2$ to 
account for the underlying 
mechanisms. By assuming that the energy dependence is dominated by the resonating final-state 
interactions, we can then write the $\pi\eta$ event distribution near the 
$K\bar{K}$ threshold as \cite{jpsiom,Oller:2000fj}
\begin{eqnarray}\label{eq.a0inv} 
\frac{d N_{\pi\eta}}{ d E_{\pi\eta}} 
= q_{\pi\eta}\, \bigg| \,c_1\, D^{-1}(s)_{\pi\eta \to \pi\eta}  
+ c_2 \, D^{-1}(s)_{ \pi\eta\to K \bar{K}}   \, \bigg|^2\,,
\end{eqnarray} 
where $q_{\pi\eta}$ denotes the three-momentum of the $\pi\eta$ system in the center-of-mass (CM) frame,  
$E_{\pi\eta} = \sqrt{s}$ is the CM energy and the matrix function $D^{-1}(s)$ is defined 
as~\cite{Oller:2000fj}   
\begin{eqnarray}
D^{-1}(s)= [1 + N^I_J(s) \cdot G(s)  ]^{-1}\,, 
\end{eqnarray} 
such that the unitarized $T$ matrix, cf. Eq.~\eqref{eq.ndt}, can be written as $\mathcal{T}=D^{-1}\cdot N$. 
In general, the parameters $c_1$ and $c_2$ in Eq.~\eqref{eq.a0inv} can be complex, but due to the 
irrelevance of an overall phase in the linear combination  of Eq.~\eqref{eq.a0inv}, just one of the 
two parameters needs to be complex. 
For definiteness, we take $c_2$ to be real in later numerical discussions and treat
$c_1$ as complex, if necessary.  Note that in  Refs.~\cite{Guo:2011pa,Guo:2012yt} it is found that 
two real parameters are  enough to reproduce the event  distribution~\footnote{Comparing with the 
formula in Refs.~\cite{Guo:2011pa,Guo:2012yt},  we use a slightly different parameterization to fit 
the $\pi\eta$ event distribution in this work.}. 

In addition, we also include the experimental $\gamma\gamma\to\pi\eta$ cross section from 
Ref.~\cite{Uehara:2009cf} in our analyses. Clearly, the strong $\pi\eta$ final-state interaction 
plays the most important role in the $\gamma\gamma\to\pi\eta$ reaction around the $a_0(980)$ resonance 
region. Based on this argument,  we use a similar expression to Eq.~\eqref{eq.a0inv} to fit 
the $\pi\eta$ cross section, with  different parameters $c_1'$ and $c_2'$, which mimic the 
$\pi\eta$ production mechanism in diphoton annihilation. The explicit formula to fit the 
$\gamma\gamma\to\pi\eta$ cross section reads
\begin{eqnarray}\label{eq.ggpeinv} 
\sigma(s) 
= \frac{\alpha^2 q_{\pi\eta}}{2s^{3/2}}\, \bigg| \,c_1' \,D^{-1}(s)_{\pi\eta \to \pi\eta}  
+ c_2'\,D^{-1}(s)_{ \pi\eta\to K \bar{K}}   \, \bigg|^2\,,
\end{eqnarray} 
with $\alpha$ the fine-structure constant. 
Analogously to  Eq.~\eqref{eq.a0inv},  just one of the parameters $c_1'$ and $c_2'$ in 
Eq.~\eqref{eq.ggpeinv} 
needs to be complex. We fix $c_2'$ as a real parameter and treat $c_1'$ as complex if required 
to improve the fit quality.

The phase shift and inelasticity can be easily read off from the $S$ matrix, which in our convention 
is related to the unitarized scattering amplitude $\mathcal{T}$ of Eq.~\eqref{eq.ndt} through 
\begin{equation}
 S  = 1 + 2 i \sqrt{\rho(s)}\cdot \mathcal{T}(s)\cdot \sqrt{\rho(s)}\,,  
\end{equation} 
with $\rho(s)=\sigma(s)/(16\pi s)$. The phase shifts $\delta_{kk}$ and $\delta_{kl}$ and 
inelasticities $\varepsilon_{kk}$ and $\varepsilon_{k l}$,  with $k\neq l$, are then given by 
\begin{align}\label{eq.defsmat}
S_{ k k} = \varepsilon_{k k} {\rm e}^{2 i \delta_{ k k}}\,, \qquad 
S_{ k l} = i \varepsilon_{k l} {\rm e}^{ i \delta_{ k l}}\,.
\end{align}

\subsection{{\boldmath$U(3)$ $\chi$}PT in a finite volume}\label{sec.introfv}

Although the experimental $\pi\eta$ event distribution~\cite{Armstrong:1991rg} and the 
$\gamma\gamma\to\pi\eta$ cross section~\cite{Uehara:2009cf} can provide some hints on 
the strong  $\pi\eta$ interactions, both of them are complicated by the complex production mechanisms 
and cannot provide direct $\pi\eta$ scattering information. In fact, direct experimental measurements 
on the $\pi\eta$ scattering, such as the phase shifts and inelasticities, are still absent. 
This is one of the key obstacles that prevents a precise determination of the $a_0(980)$ properties. 

Recently,  the first calculation of $\pi\eta$ scattering, including the $K\bar{K}$ and $\pi\eta'$ 
coupled channels, has been carried out in lattice QCD~\cite{Dudek:2016cru}. The simulations are done 
with three different lattice volumes, but only one large pion mass ($m_\pi \sim 391$~MeV) is used. 
By performing the analysis in many moving frames, a large number of discrete energy levels in  three 
volumes are obtained. The rich spectra in a finite box contain direct information on the $\pi\eta$ 
scattering. In Ref.~\cite{Dudek:2016cru}, a large number of different $K$-matrix parametrizations are
used to extract the phase shifts and inelasticities from the various finite-volume  
energy levels. In this work, we propose to use another framework, the unitarized $U(3)$ $\chi$PT, 
to reanalyze the discrete spectra.

In order to use this approach to describe the lattice energy levels, we first 
need to include the finite-volume effects in unitarized $U(3)$ $\chi$PT. Generally speaking, 
there are two different 
kinds of volume dependence of the scattering amplitudes. First, there are the contributions which are
exponentially suppressed $\propto \exp(-m_P L)$, where $m_P$ is the mass of 
the lightest particle in the problem at hand and $L$ denotes the size of the box. 
Second, if the energy is above threshold,  there are contributions that are only 
power suppressed $\propto 1/L^3$ and behave irregularly. It can be demonstrated 
(see, e.g., Refs.~\cite{Bedaque:2006yi,Albaladejo:2012jr}) 
that only the $s$-channel contributions can lead to the power-law corrections, 
while the crossed channels  give rise only to the exponentially suppressed terms 
(there are exponentially suppressed $s$-channel contributions as well). 
This indicates that the power-suppressed contributions in the unitarized chiral amplitude 
in Eq.~\eqref{eq.ndt} are generated solely by the modification of the function $G(s)$, which 
incorporates the $s$-channel unitarity cut. On the contrary, 
the function $N(s)$, which contains the crossed-channel contributions by construction, 
contributes to the exponentially suppressed volume dependences only. In the present work, we include 
the important finite-volume effects through the function $G(s)$ in Eq.~\eqref{eq.ndt} and neglect 
the exponentially suppressed volume dependence of the $N(s)$ function. The same prescription has 
been used in the previous studies within the same  
framework~\cite{Doring:2011vk,Doring:2012eu,Doring:2013glu,Doring:2011nd,Xie:2012np,Chen:2012rp,Zhou:2014ana,Geng:2015yta,Albaladejo:2016jsg,Lu:2016kxm,MartinezTorres:2011pr}.

Furthermore, we would like to comment on the relation of  unitarized $\chi$PT 
in a finite volume with the L\"uscher approach~\cite{luescher-torus}. 
In fact, these two approaches are quite similar since, as can be easily shown, the finite-volume 
modification of the function $G(s)$ can be expressed through the L\"uscher zeta function 
up to the exponentially suppressed contributions~\cite{Doring:2011vk}. Thus, the only difference 
with the L\"uscher approach amounts to the use of the different $K$-matrix parametrizations 
in the infinite volume: In the unitarized $\chi$PT case one effectively parameterizes the $K$ matrix 
through the solution of the coupled-channel equations, whereas simple algebraic parametrizations were 
used in Refs.~\cite{Wilson,Dudek:2016cru,Prelovsek}. If $L$ becomes smaller, the exponentially 
suppressed terms become important and these two approaches are no longer equivalent. However, we 
do not consider this case here.

Following the prescription in Ref.~\cite{Doring:2011vk}, the finite-volume effects can be implemented 
in the two-point loop function $G(s)$ in Eq.~\eqref{eq.defg} by replacing the continuous three-momentum 
integral with the sum of allowed discrete momenta in the finite box with  periodic boundary conditions. 
In order to perform the sum, it is convenient to integrate out the zeroth component of the 
four-momentum integral in  Eq.~\eqref{eq.defg}. This gives 
\begin{eqnarray}\label{eq.defg3d}
 G(s)^{\rm cutoff}= \int^{|\vec{q}|<q_{\rm max}} \frac{{\rm d}^3 \vec{q}}{(2\pi)^3} \, I(|\vec{q}|) \,,
\end{eqnarray}
where 
\begin{eqnarray}
I(|\vec{q}|) &=& \frac{w_1+w_2}{2w_1 w_2 \,[E^2-(w_1+w_2)^2]}\,, \\
w_i &=&\sqrt{|\vec{q}|^2+m_i^2} \,,  \quad s=E^2 \,,  
\end{eqnarray} 
and the ultraviolet three-momentum cutoff $q_{\rm max}$ is introduced to regularize the divergent integral.
One could also use other regularization methods to obtain finite results, such as the dimensional 
regularization, cf.~Eq.~\eqref{eq.gfuncdr}, or include different types of form  
factors~\cite{Doring:2011vk}.
We use sharp cutoffs below. When calculating the function  $G(s)$ in a finite box of length $L$ 
with periodic boundary conditions, one obtains 
\begin{eqnarray}
 \widetilde{G}= \frac{1}{L^3} \sum_{\vec{n}}^{|\vec{q}|<q_{\rm max}} I(|\vec{q}|)\,,
\end{eqnarray}
where a tilde on top of a symbol is introduced to denote the quantities in the finite volume and 
\begin{eqnarray}
 \vec{q}&=& \frac{2\pi}{L} \vec{n}, \quad  \vec{n} \in \mathbb{Z}^3 \,.
\end{eqnarray}
The difference between the infinite- and finite-volume  functions
can then be calculated through 
\begin{eqnarray}
 \Delta G &=& \widetilde{G} - G^{\rm cutoff} \nonumber \\
  & = & \frac{1}{L^3} \sum_{\vec{n}}^{|\vec{q}|<q_{\rm max}} I(|\vec{q}|) -   \int^{|\vec{q}|<q_{\rm max}} 
\frac{{\rm d}^3 \vec{q}}{(2\pi)^3} I(|\vec{q}|)\,. 
\end{eqnarray}
Note that the finite-volume correction $\Delta G$ is independent of the cutoff $q_{\rm max}$ in the 
limit $L\to\infty$ due to the cancellation of the cutoff dependences in the two terms in 
this  equation, as explicitly demonstrated in Ref.~\cite{Doring:2011vk}. In the 
practical calculation, we have explicitly verified that the cutoff dependence of $\Delta G$ 
is indeed quite small. In general, taking $q_{\rm max}=\frac{2\pi}{L}n_{\rm max}$ and $L$ around 2~fm 
($m_\pi L \sim 4$), 
the change of $\Delta G$ for the $\pi\eta$ channel is typically smaller than one percent when 
increasing $n_{\rm max}$ from 20 to 30.

One can then add the finite-volume correction $\Delta G$ to the infinite-volume result 
$G^{\rm DR}$ in Eq.~\eqref{eq.gfuncdr} to get the final expression of the $G$ function 
used in our study 
\begin{eqnarray}\label{eq.gfuncfvdr}
 \widetilde{G}^{\rm DR}= G^{\rm DR} + \Delta G \,. 
\end{eqnarray}
This is the prescription followed in Ref.~\cite{MartinezTorres:2011pr}. 
The expression $\widetilde{G}^{\rm DR}$ evaluated in the finite box should 
always be real in the whole energy region, which is guaranteed in 
Eq.~\eqref{eq.gfuncfvdr} due to the cancellation of the imaginary parts in $G^{\rm DR}$ and 
the cutoff integral in $\Delta G$ above threshold.

The two-point loop function $G(s)$ in Eq.~\eqref{eq.defg} is manifestly Lorentz invariant
in the infinite volume. 
However, this is not the case for the finite-volume situation, where the Lorentz invariance is lost.
One then needs to work out the explicit form of the loop function, 
when boosting from one frame to another. 
This issue has been addressed in Refs.~\cite{Doring:2012eu,Gockeler:2012yj,Roca:2012rx} 
and we briefly recapitulate
the main results in order to introduce the necessary notations.

In the CM frame of the two particles,  one has $\vec{q}_1^{\,*}=-\vec{q}_2^{\,*}$, where 
we follow the convention that any quantity defined in the CM frame is marked with an asterisk. 
Now let us consider the two-particle system in a moving frame with  total four-momentum 
$P^\mu=(P^0, \vec{P})$. The square of the CM energy of the two-particle system is then given 
by $s=E^2=(P^0)^2-|\vec{P}|^2$. The three-momenta of the two particles in the moving frame 
are $\vec{q}_1$ and $\vec{q}_2=\vec{P}-\vec{q}_1$. Boosting to the CM frame, i.e., transforming 
$\vec{q}_{i=1,2}$ to $\vec{q}_{i=1,2}^{\,*}$, one straightforwardly obtains 
\begin{eqnarray}\label{eq.lrtr1}
 \vec{q}_i^{\,*}= \vec{q}_i + \bigg[ \bigg(\frac{P^0}{E} - 1\bigg)\frac{\vec{q}_i\cdot \vec{P}}{|\vec{P}|^2} - \frac{q_i^{\,0}}{E} \bigg]\vec{P}\,. 
\end{eqnarray} 
Furthermore, following Ref.~\cite{Doring:2012eu}, we notice that one is free to impose the on-shell 
relation between energy and  three-momentum: 
$q_i^{*\,0}=\sqrt{|\vec{q}_i^{\,*}|^2+m_i^2}$. 
We also mention that this is equivalent to enforcing the on-shell condition 
for $q_i^{\,0}=\sqrt{|\vec{q}_i|^2+m_i^2}$, 
which automatically leads to the on-shell condition for $q_i^{*\,0}$ through the following 
Lorentz transformation: 
\begin{eqnarray}\label{eq.q0}
 q_i^{\,0}= \frac{q_i^{*\,0} E + \vec{q}_i\cdot\vec{P}}{P^0} \,.
\end{eqnarray}
In order to establish the relation of the functions $G(s)$ in the moving and CM frames, 
one needs to calculate the Jacobian of the transformation from 
${\rm d}^3 \vec{q}_i^{\,*}$ to ${\rm d}^3 \vec{q}_i$. In this respect, 
it is convenient to rewrite Eq.~\eqref{eq.lrtr1}, substituting Eq.~\eqref{eq.q0}. This gives 
\begin{eqnarray}\label{eq.lrtr2}
 \vec{q}_i^{\,*}= \vec{q}_i + \bigg[ \bigg(\frac{E}{P^0} - 1\bigg)\frac{\vec{q}_i\cdot \vec{P}}{|\vec{P}|^2} 
- \frac{q_i^{*\,0}}{P^0} \bigg]\vec{P}\,, 
\end{eqnarray} 
where according to the on-shell condition for $q_{i=1,2}^{*\,0}$ one has 
\begin{eqnarray}
 q_1^{*\,0}= \frac{E^2+m_1^2-m_2^2}{2E} \,, \quad  q_2^{*\,0}= \frac{E^2+m_2^2-m_1^2}{2E} \,.
\end{eqnarray}
Using Eq.~\eqref{eq.lrtr2} it is straightforward to obtain the Jacobian 
\begin{eqnarray}
 \int {\rm d}^3{\vec{q}}_1^{\,*} =  \frac{E}{P^0} \int {\rm d}^3 \vec{q}_1\,. 
\end{eqnarray} 
Then, the integral can be discretized through the following substitution  
\begin{eqnarray}
\int^{|\vec{q}_1|^{\,*}<q_{\rm max}} \frac{{\rm d}^3{\vec{q}_1}^{\,*}} {(2\pi)^3} I(|\vec{q}_1^{\,*}|) 
\,\, \Longrightarrow  \nonumber \\ 
\widetilde{G}^{\rm MV}= \frac{E}{P^0 L^3}  \sum_{\vec{q}_1}^{|\vec{q}_1^{\,*}|<q_{\rm max}} I(|\vec{q}_1^{\,*}(\vec{q}_1)|)\,,
\label{eq.gfuncfvmv}
  \end{eqnarray} 
with 
\begin{eqnarray}
\vec{q}_1=\frac{2\pi}{L}\vec{n}\,, \quad \vec{n} \in \mathbb{Z}^3\,, \\ 
\quad \vec{P}=\frac{2\pi}{L}\vec{N}\,,  \quad  \vec{N} \in \mathbb{Z}^3\,. \label{eq.pfv}
\end{eqnarray} 
Note that the CM three-momentum of the two-particle system $\vec{P}$ in the finite box should 
only take the discrete values shown in Eq.~\eqref{eq.pfv} in order to impose the  condition 
$\vec{q}_1+\vec{q}_2=\vec{P}$. The final $G$ function after taking into account the 
finite-volume corrections in the moving frame takes the form 
\begin{eqnarray}\label{eq.gdrfvmv}
 \widetilde{G}^{\rm DR, MV}= G^{\rm DR} + \Delta G^{\rm MV}  \,,
\end{eqnarray}
where
\begin{equation}
\Delta G^{\rm MV} =   \widetilde{G}^{\rm MV} - G^{\rm cutoff}\,,
\end{equation}
with $G^{\rm DR}, G^{\rm cutoff}$ and $\widetilde{G}^{\rm MV}$ given in 
Eqs.~\eqref{eq.gfuncdr}, \eqref{eq.defg3d} and \eqref{eq.gfuncfvmv}, respectively.

Before ending this section, we briefly comment  on the partial wave mixing effects for a  
nonvanishing total momentum $\vec{P}$ in the finite box. A noticeable 
difference between $\vec{P}=0$ (CM frame) and $\vec{P}\neq 0$ (moving frame) is that, in the former 
case, the $S$ wave can mix with the $G$ wave only (the effect of such a mixing is presumed 
to be tiny), whereas in
the latter case, there are more mixing patterns:  Even the mixing of the 
$S$ and $P$ waves can not be excluded in general. 
The mixing terms between different partial waves could give some visible effects for some specific 
channels, such as the $\pi K$ $S$- and $P$-wave scattering, while in some other cases the 
mixing effects are tiny, such as the $\pi\pi$ $S$- and $D$-wave scattering~\cite{Doring:2012eu}. 
Due to the fact that the isospin for the $P$-wave $\pi\pi$ is 1 and the isospin for $S$-wave 
$\pi\pi$ is 0 or 2, there is no mixing between $\pi\pi$ $S$- and $P$-wave amplitudes. 

The situation in $\pi\eta^{(')}$ and $K\bar{K}$ scattering is more subtle. 
The $G$ parity of $\pi\eta^{(')}$ scattering is definite and negative. There is no $P$-wave or higher odd 
waves in $K\bar{K}$ scattering with negative $G$ parity. Only even-wave $K\bar{K}$ scattering, such 
as $S$ and $D$ waves, can have negative $G$ parity.  For the $P$-wave  $\pi\eta^{(')}$ scattering, 
one has the $J^{PC}=1^{-+}$ 
exotic quantum numbers and therefore one does not expect any strong interactions in the low energy 
region~\footnote{It is also very unlikely that the possible exotic states $\pi_1(1400)$ and 
$\pi_1(1600)$~\cite{Olive:2016xmw} will have important impact around the $K\bar{K}$ threshold region.}. 
As for the $D$-wave $\pi\eta^{(')}$ scattering, it only starts to become important around 
the $a_2(1320)$  region, and it shows very little impact near the 
$K\bar{K}$ threshold,  which is explicitly verified in the lattice simulations in Ref.~\cite{Dudek:2016cru}. 
Based on these arguments, it seems quite plausible that the mixing effects between the higher partial waves 
and the $S$ wave in $\pi\eta$, $K\bar{K}$ and $\pi\eta'$ scattering are small. 
Therefore, in the present study we neglect the higher partial wave effects, 
which is explicitly verified to be a good assumption in Ref.~\cite{Dudek:2016cru}.  

In summary the formulas that we use to determine the lattice finite-volume energy levels are 
\begin{equation}\label{eq.detmv} 
 \det[I+ N^{1}_{0}(s)\cdot\widetilde{G}^{\rm DR,MV}] = 0\,,
\end{equation}
for the moving frames, and 
\begin{equation}\label{eq.detcm}
 \det[I+ N^{1}_{0}(s)\cdot \widetilde{G}^{\rm DR}] = 0\,,
\end{equation}
for the CM frame. 
The matrix $I$ in the previous two equations denotes the $3\times 3$ unit matrix,
and $\widetilde{G}^{\rm DR,MV}$ and $\widetilde{G}^{\rm DR}$ should be understood as $3\times 3$ 
diagonal matrices,
with their matrix elements calculated for the $\pi\eta$, $K\bar{K}$ and $\pi\eta'$ channels.

\section{Global fits to the lattice energy levels and experimental data}\label{sec.fit}

In this section, we discuss the global fits to the lattice finite-volume energy levels and 
the experimental data, including a  $\pi\eta$ event distribution~\cite{Armstrong:1991rg} and 
the $\gamma\gamma\to\pi\eta$ cross section~\cite{Uehara:2009cf}. 
On the one hand, the lattice energy levels contain the direct $\pi\eta$ scattering information, but 
the numerical simulations are done with a relatively heavy pion mass around $391$~MeV. 
On the other hand, the experimental data encode the $\pi\eta$ dynamics at physical masses, 
but both the event distribution and cross section of the diphoton fusion are affected 
by the complex production mechanisms, which usually bring additional uncertainties 
when extracting the direct $\pi\eta$ scattering information. 
Nonetheless, it is clear that the global fits to both kinds of data from lattice and 
experiment impose  stronger constraints on the $\pi\eta$ scattering amplitudes than the fit 
to only one set of these data.

Concerning the lattice simulations, we focus on the energy levels below the $\pi\eta'$ 
threshold and the data points considered in our fits are explicitly shown in Figs.~\ref{fig.lolev000} 
and \ref{fig.lolev001}. This amounts to 47 data points which are provided by the authors of  
Ref.~\cite{Dudek:2016cru} with the correlation information for those obtained within the same 
lattice volume. The  data points considered in our work are exactly the same as those 
fitted in the two-channel formalism by using the various two-channel $K$-matrix parametrizations 
in Ref.~\cite{Dudek:2016cru}. For the $\pi\eta$ event  distribution, there are 11 data 
points~\cite{Armstrong:1991rg}, 
which are shown in Fig.~\ref{fig.eventdis}. Note that the background parts given in 
Ref.~\cite{Armstrong:1991rg} are explicitly extracted when we fit the event distribution. 
The $\gamma\gamma\to\pi\eta$ cross section points~\cite{Uehara:2009cf} 
are shown in Fig.~\ref{fig.crosssection}, which amounts to 10 more data points. 
The systematic error bands given  in Ref.~\cite{Uehara:2009cf} are taken into account in the fits.

For the fits to lattice energy levels, we take the masses for $\pi, K, \eta$ and $\eta'$ 
from  Ref.~\cite{Dudek:2016cru} 
\begin{equation}\label{eq.masslat}
 m_\pi=391.3\pm0.7~\mathrm{MeV}\,, \,\,  m_K=549.5\pm0.5~\mathrm{MeV}\,, \,\,  
m_\eta=587.2\pm1.1~\mathrm{MeV}\,, \,\,  m_{\eta'}=929.8\pm5.7~\mathrm{MeV}\,.
\end{equation}
For the fits to experimental data the values of the masses are the same as in Ref.~\cite{Guo:2011pa}
\begin{equation}\label{eq.massphy}
 m_\pi= 137.3~\mathrm{MeV}\,, \quad  m_K= 495.6~\mathrm{MeV}\,, \quad  m_\eta= 547.9~\mathrm{MeV}\,, 
\quad  m_{\eta'}= 957.7~\mathrm{MeV}\,.
\end{equation}

The $\eta$-$\eta'$ mixing angle in the flavor octet-singlet basis is needed in our theoretical model, 
as can be seen from Eq.~\eqref{eq.pwlo}. In order to calculate the LO $\eta$-$\eta'$ mixing angle in 
Eq.~\eqref{eq.loangle}, we first need to know the LO masses for the pion and kaon, i.e., $\overline{m}_\pi$ 
and $\overline{m}_K$. In the Appendix 
of Ref.~\cite{Guo:2011pa}, the explicit formulas are provided to calculate these two quantities and 
we do not quote the expressions here. Using the masses from the lattice simulations in 
Eq.~\eqref{eq.masslat}, the LO $\eta$-$\eta'$ mixing angle turns out to be 
\begin{equation}
\theta=(-10.0 \pm 0.1)^{\circ}\,, 
\end{equation} 
which is in good agreement with the values given in Refs.~\cite{Dudek:2011tt,Dudek:2013yja} and 
can be compared with  the value $\theta^{\rm phys}=-16.2^{\circ}$ at the physical masses~\cite{Guo:2011pa}. 
Note that this mixing angle has also recently been calculated by using different lattice actions: 
see, e.g., Ref.~\cite{Urbach}.

Another important quantity that is needed in our calculation, as can be seen in Eq.~\eqref{eq.pwlo}, 
is the pion decay constant $F_\pi$. Its value at the specific masses of Eq.~\eqref{eq.masslat}  
is not reported in Ref.~\cite{Dudek:2016cru}. Therefore, we need to estimate $F_\pi$ at the 
unphysical masses within our approach. The one-loop $U(3)$ $\chi$PT result is already 
given in Ref.~\cite{Guo:2011pa}, 
which reads 
\begin{eqnarray}
\label{eq.ffpi}
F_\pi &=& F \bigg\{ 1 -
\frac{1}{16\pi^2 F^2}\bigg[   m_\pi^2 \ln{\frac{m_\pi^2}{\mu^2}} + \frac{m_K^2}{2} \ln{\frac{m_K^2}{\mu^2}}  \bigg]
\nonumber \\ &&
\qquad\qquad
+\bigg[ \frac{4 \widetilde{c}_d \,\widetilde{c}_m (m_\pi^2+2m_K^2)}{F^2 M_{S_1}^2}
- \frac{8 c_d\, c_m\, (m_K^2-m_\pi^2)}{3 F^2 M_{S_8}^2} \bigg] \bigg\}  \,.
\end{eqnarray}
In this equation, $\widetilde{c}_{m,d}$ and $c_{m,d}$ are the couplings of the $SU(3)$ singlet and 
octet bare  scalar resonances with  masses $M_{S_1}$ and $M_{S_8}$, which were introduced at the 
Lagrangian level. 
We take their values as determined in Ref.~\cite{Guo:2011pa}.

We point out that up to one-loop level precision there is an ambiguity in choosing the pion decay 
constant appearing inside the curly  brackets on the right hand side (rhs) of Eq.~\eqref{eq.ffpi}. 
For example, one can also use  the renormalized $F_\pi$ inside the curly brackets 
in this equation, since for a one-loop  calculation the difference is of higher order.
In order to conveniently deal with this ambiguity, 
we impose two extra conditions to determine the expression for $F_\pi$. 
The first condition is that  one should recover the physical value of $F_\pi=92.4$~MeV 
when using the physical pion and kaon masses with a proper value of $F$. 
The other condition is that in the meantime we require that our extrapolation formula 
for $F_\pi$ reproduces other existing lattice simulation  
results~\cite{Davies:2003fw,Davies:2003ik,Aoki:2010dy,Arthur:2012yc}, 
which were analyzed in Ref.~\cite{Guo:2014yva} in a chiral framework, when using the 
specific masses in Eq.~\eqref{eq.masslat}. Guided by these requirements, we find that when  
substituting $F=77.0$~MeV in Eq.~\eqref{eq.ffpi}, we get the correct value for $F_\pi$ 
with physical pion and kaon masses, while  
taking the masses in Eq.~\eqref{eq.masslat} leads to $F_\pi=105.9$~MeV, a value that is 
reasonably close to other lattice simulation 
results~\cite{Davies:2003fw,Davies:2003ik,Aoki:2010dy,Arthur:2012yc}. Therefore, we take 
$F_\pi=105.9$~MeV as the central value in our fits to the lattice energy levels of 
Ref.~\cite{Dudek:2016cru}.

However, in order to make a further test about the influence of using different $F_\pi$ extrapolation 
forms on the extracted energy levels, we also replace $F$ inside the curly brackets 
in Eq.~\eqref{eq.ffpi} with the physical value of $F_\pi$. In this case we find that with $F=81.1$~MeV 
the rhs of Eq.~\eqref{eq.ffpi} leads to $F_\pi=92.4$~MeV at physical pion and kaon masses, 
and then it predicts $F_\pi=102.3$~MeV with the masses in Eq.~\eqref{eq.masslat}. We consider 
the differences of $F_\pi$ obtained with the two different extrapolation forms as an additional 
source of uncertainty in our study, which can be treated as a 
systematic error. In summary at the masses given in Eq.~\eqref{eq.masslat} we use 
\begin{equation}\label{eq.fpilaterr} 
 F_\pi = 105.9 \pm 3.6~\,{\rm MeV}
\end{equation}
to extract the finite-volume energy levels. Comparing with the 
lattice results given in Refs.~\cite{Davies:2003fw,Davies:2003ik,Aoki:2010dy,Arthur:2012yc},
we may conclude that our estimate of the error in $F_\pi$, Eq.\eqref{eq.fpilaterr}, is quite conservative.
When fitting to experimental data, we always fix $F_\pi$ at its physical value.

\subsection{Leading-order fit}\label{sec.lofit}

In this part we present the LO fit results. The $N^{1}_0(s)$ matrix function in 
Eq.~\eqref{eq.defnfunction} is simply given by the LO $T^{1}_0(s)^{(2)}$ in Eq.~\eqref{eq.pwlo}. 
At this order the only unknown parameters in the unitarized chiral amplitudes are the three 
subtraction constants $a_{\pi\eta}, a_{K\bar{K}}$ and $a_{\pi\eta'}$. 
The fits turn out to be rather insensitive to the value of  $a_{\pi\eta'}$ (a feature that is also seen 
in the NLO fits discussed in the next section). 
Therefore we will always fix its value to be equal to $a_{K\bar{K}}$, both in the LO and NLO 
fits. Furthermore in the LO fit we find that just one common subtraction constant for the three channels 
is already enough to obtain a good fit. Leaving the value for $a_{K\bar{K}}$ free barely improves 
the fit quality. 
Therefore we impose $a_{\pi\eta}=a_{K\bar{K}}=a_{\pi\eta'}$ for this case. 
As discussed in Sec.~\ref{sec.introu3xpt} we need to include additional parameters in order to describe 
a $\pi\eta$ event distribution and the $\gamma\gamma\to\pi\eta$ cross section, cf. Eqs.~\eqref{eq.a0inv}  
and \eqref{eq.ggpeinv}.
For the $\pi\eta$ event distribution, two real parameters $c_1$ and $c_2$ are found to be enough 
to reproduce the data well.  For the $\gamma\gamma\to\pi\eta$ cross section, we find that just 
one real parameter $c_2'$ alone is able to give reasonable description of the experimental data, 
and take $c_1'=0$.

\begin{figure}[htbp]
 \centering
\includegraphics[width=0.7\textwidth,angle=-0]{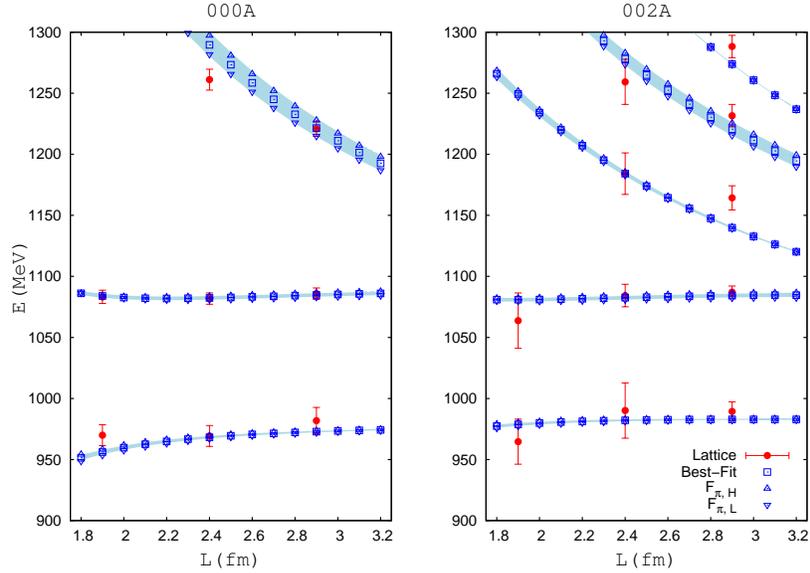} 
\caption{Fit results of the finite-volume energy levels for the ensembles $000A$ and $002A$ at 
leading order. The lattice data are taken from Ref.~\cite{Dudek:2016cru}. The square symbols 
represent the results from our best fit and the shaded areas correspond to the 1-$\sigma$ uncertainties.  
The upwards and downwards  
triangle symbols, labeled as $F_{\pi, {\rm H} }$ and $F_{\pi, {\rm L}}$, denote the results obtained 
by taking the upper and lower limits of $F_\pi$ in Eq.~\eqref{eq.fpilaterr}. }
   \label{fig.lolev000}
\end{figure}

The LO fit gives a reasonable description of the overall data, with a $\chi^2/{\rm d.o.f}
=104.5/(68-4)\simeq 1.69$. 
The chi square contributed by the 47 lattice energy levels is 90.2, and the chi square from the 21 
experimental data  is 14.3. Note that the correlation information among the lattice energy 
levels within the same volume~\cite{Dudek:2016cru} is considered in our fit. The value of the  
subtraction constant  from this fit is 
\begin{equation}\label{eq.lofit}
 a_{\pi\eta}= -1.44\pm0.15\pm0.01\,. 
\end{equation}
The values for the phenomenologically motivated parameters $c_1, c_2$ and $c_2'$ are  
$c_1=0.44\pm0.14\pm0.00$~MeV${^{-1}}$, $c_2=-0.27\pm0.11\pm0.00$~MeV${^{-1}}$ and $c_2'=2.18\pm0.36\pm0.02$. 
The first error bar of each parameter corresponds to the statistical one and the second one
is caused by the uncertainties of the unphysical masses in Eq.~\eqref{eq.masslat}. Note that 
when using  Eq.~\eqref{eq.ggpeinv} to fit the $\gamma\gamma\to\pi\eta$ 
cross section, we have introduced the proper normalization factor to transform the 
unit MeV$^{-2}$ to nanobarn. 
The statistical error bars of the parameters $a_{\pi\eta}$, $c_1,c_2$ and $c_2'$ are calculated in the 
following way. We randomly vary the parameters around their central values from the best fit, 
recalculate the corresponding new chi square and then only keep the ones that give $\chi^2\leq 
\chi_0^2 + \sqrt{2\chi_0^2}$ (with $\chi_0^2$ the chi-square 
value from the best fit), i.e. those within the 1-$\sigma$ standard deviation. 
In order to estimate the influences on the parameters from the uncertainties of the unphysical 
masses in Eq.~\eqref{eq.masslat},  we have performed a large number of fits by randomly 
varying the masses within uncertainties. It turns out that the variances of the central 
values of the fitted parameters are one order smaller than the statistical error bars and hence 
negligible. With those parameter configurations within 1-$\sigma$ uncertainty, we also 
calculate the error bands of the other quantities,  including the finite-volume energy 
levels, event distribution, 
cross section, phase shifts, inelasticities, the pole positions, and corresponding residues.

\begin{figure}[htb]
   \centering
   \includegraphics[width=0.9\textwidth,angle=-0]{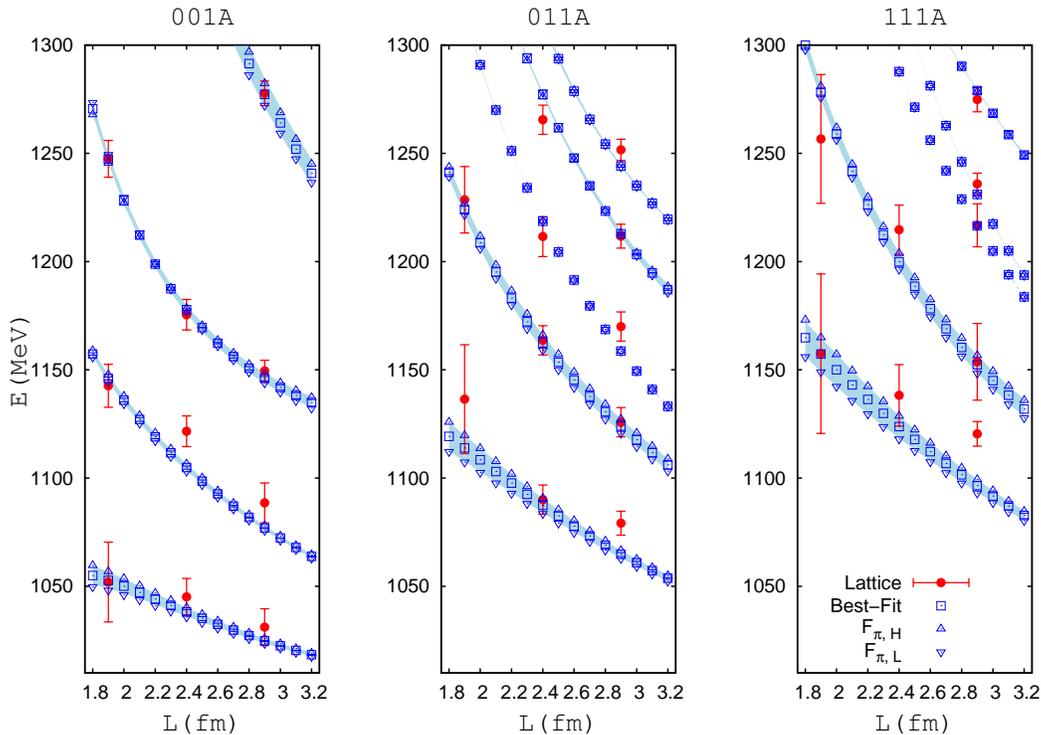} 
  \caption{Fit results of the finite-volume energy levels for  the ensembles $001A$, $011A$ and $111A$ 
           at leading order. The lattice  data are taken from Ref.~\cite{Dudek:2016cru}. For notations, 
           see Fig.~\ref{fig.lolev000}.  }
   \label{fig.lolev001}
\end{figure}

\begin{figure}[htbp]
   \centering
   \includegraphics[width=0.8\textwidth,angle=-0]{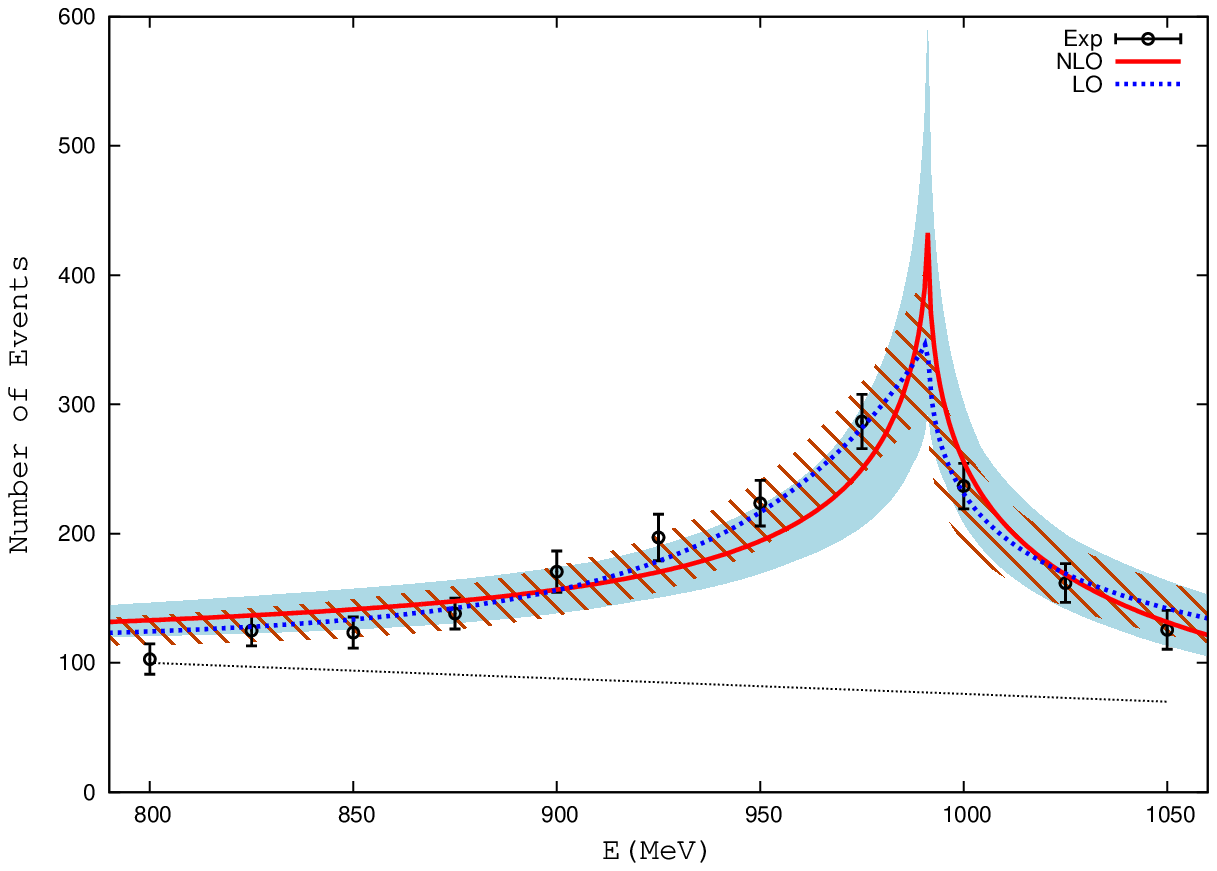} 
  \caption{The $\pi\eta$ event distribution. The experimental data are taken from 
          Ref.~\cite{Armstrong:1991rg}. 
           The black dotted line represents the background extracted from this reference. The blue dotted 
           and solid lines denote the LO and NLO fit results, respectively. The shaded areas correspond 
           to the 
           1 $\sigma$ error bands.   }
   \label{fig.eventdis}
\end{figure}

\begin{figure}[htbp]
   \centering
   \includegraphics[width=0.8\textwidth,angle=-0]{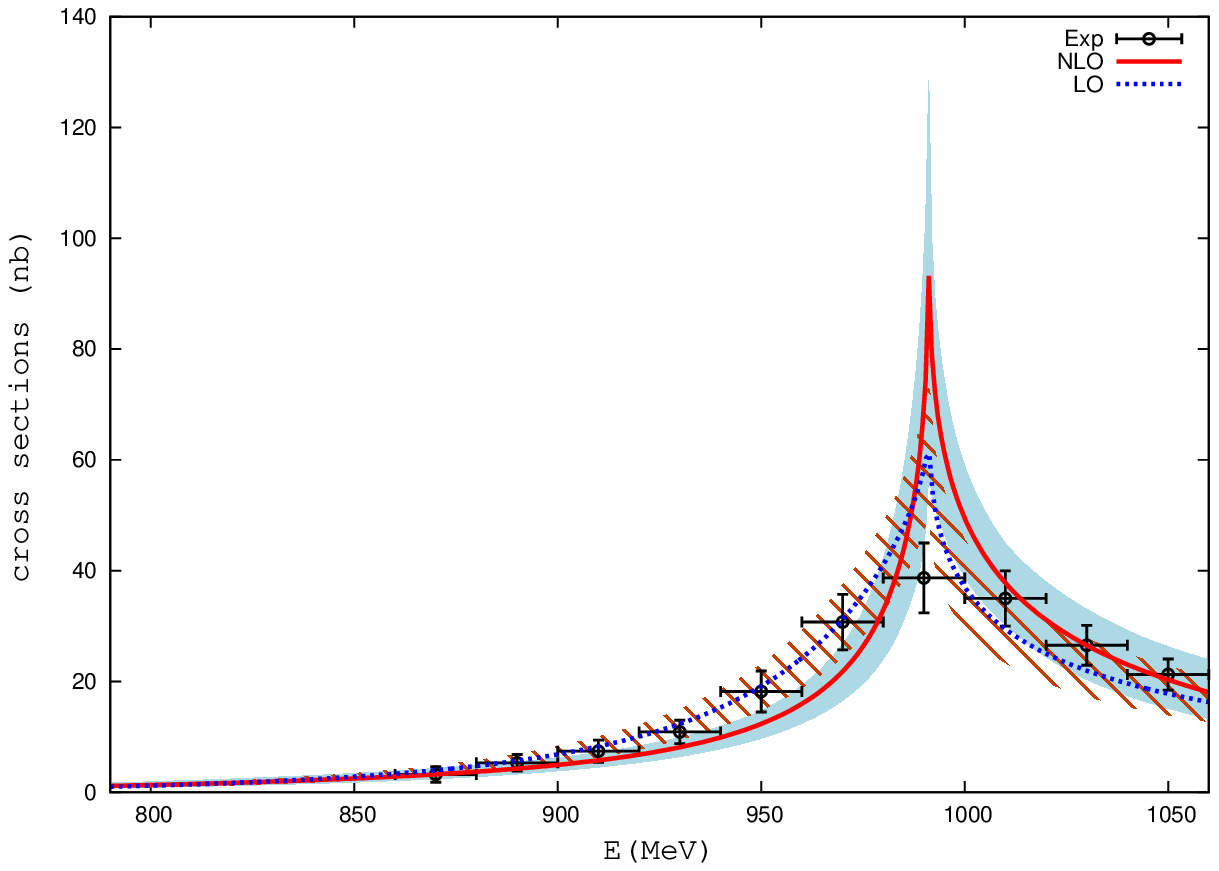} 
  \caption{Cross section of $\gamma\gamma\to \pi\eta$. The experiment data are taken from 
           Ref.~\cite{Uehara:2009cf} and the systematic error bands from this reference are included 
           in our fits. For notations, see Fig.~\ref{fig.eventdis}.  }
   \label{fig.crosssection}
\end{figure}

The reproduction of the lattice energy levels is shown in Figs.~\ref{fig.lolev000} and \ref{fig.lolev001}, 
where the square symbols stand for the results from our best fit and the shaded areas correspond to 
the 1 $\sigma$ error bands. The upwards and downwards triangle symbols denote the results calculated 
for the upper and lower limits of $F_\pi$ in Eq.~\eqref{eq.fpilaterr}. The fit results for the 
$\pi\eta$ event distribution and 
$\gamma\gamma\to\pi\eta$ cross section are given in Figs.~\ref{fig.eventdis} and \ref{fig.crosssection}, 
respectively. The LO best fits are plotted in blue by the dotted lines, and their hatched 
surrounding  areas present the 1 $\sigma$ uncertainties as explained before.

It is interesting to discuss two different variants of LO fits. 
In one case, we study the influences of using different pNGB decay constants in the scattering 
amplitudes in  Eq.~\eqref{eq.pwlo}. Although distinguishing different pNGB decay constants is beyond 
the LO accuracy, it may cause a visible effect when performing the chiral extrapolation. Here we 
do a tentative study of this effect by only distinguishing the kaon decay constant $F_K$ from 
the others, due to the prominent $K\bar{K}$ threshold enhancement around the $a_0(980)$ energy 
region. To be more specific, we replace one $F_\pi$ in the $\pi\eta\to K\bar{K}$, 
$K\bar{K}\to\pi\eta'$ amplitudes in Eq.~\eqref{eq.pwlo} by $F_K$ and replace $F_\pi^2$ in 
$K\bar{K}\to K\bar{K}$ by $F_K^2$. At the physical mass, $F_K$ is set to  110.1~MeV~\cite{Olive:2016xmw} 
and when taking the unphysical masses in Eq.~\eqref{eq.masslat} $F_K$ is  fixed to 115.0~MeV, a 
value consistent with the previous lattice  
determinations~\cite{Davies:2003fw,Davies:2003ik,Aoki:2010dy,Arthur:2012yc}. We mention that 
using a somewhat different value for $F_K$ at the unphysical meson masses only moderately changes 
our discussion below, since its effect can be compensated by slightly adjusting the subtraction constants. 
The resulting parameters from  the fit are $a_{\pi\eta}=-1.73\pm 0.16$, $c_1=0.31\pm0.14$, 
$c_2=-0.32\pm0.11$ and $c_1'=2.13\pm0.36$, with $\chi^2/d.o.f=107.1/(68-4)$.  Even though the subtraction 
constant is decreased by about 20\% due to the replacement of $F_\pi$ by $F_K$ in the amplitudes 
involving kaon, we will show in Secs.~\ref{sec.latdiscuss} and \ref{sec.phydiscuss} that the fits 
using different meson decay constants lead to qualitatively similar phase shifts and inelasticities 
to the case with a common $F_\pi$ in all amplitudes. 
In the other case, we perform the fit by only including the lattice energy levels. The resulting 
parameter turns out to be  $a_{\pi\eta}=-1.49\pm0.20$, with $\chi^2/d.o.f=89.3/(47-1)$. The value 
obtained here is perfectly compatible with the result in Eq.~\eqref{eq.lofit} from the global fit 
by simultaneously including both the lattice and experimental data. In other words, the LO 
expression can give a consistent descriptions of both the lattice data with large unphysical masses 
and the experimental data with physical masses.

\subsection{Next-to-leading-order fit}\label{sec.nlofit}

As demonstrated in the previous section and also in many earlier
papers~\cite{Oller:1997ti,Oller:gamma,Oller:1998hw,Oller:1998zr,Oller:2002na},  
the LO unitarized chiral amplitudes can already reasonably describe the various $\pi\eta$ reactions 
around the $K\bar{K}$ threshold energy region. As a result, it is reasonable to require that 
including the higher order effects in the unitarized amplitudes should not spoil the LO results.
Therefore, as a first step to perform the NLO fits, we impose the condition that the NLO 
unitarized chiral amplitudes stay close to the LO results within a $20\%$ uncertainty around 
the $K\bar{K}$ threshold. 
This condition -- in addition to fitting the lattice energy levels, $\pi\eta$ event distribution and 
the $\gamma\gamma\to\pi\eta$ cross section -- stabilizes the  fit, given the numerous free parameters. 
After obtaining  good fits, we finally release the closeness condition of the LO and NLO amplitudes. 
We find that in this way the fit is stable, and the final NLO amplitudes still 
qualitatively resemble the LO ones.

There are more parameters in the NLO unitarized chiral amplitudes than in the LO ones. We fit 
the three subtraction constants $a_{\pi\eta}, a_{K\bar{K}}$ and $a_{\pi\eta'}$, which  appear in the 
$\pi\eta$, $K\bar{K}$ and $\pi \eta^\prime$ channels. The other parameters are already well 
determined in Ref.~\cite{Guo:2011pa} and we take the values therein. 
At NLO, we find that it is impossible to obtain a good fit with just one subtraction constant. 
Both $a_{\pi\eta}$ and $a_{K\bar{K}}$ are fitted in this case, while the fits are quite insensitive 
to $a_{\pi\eta'}$; thus, we simply fix its value  to the one of $a_{K\bar{K}}$. 
For the additional parameters mimicking  the $\pi\eta$ production mechanisms in Eqs.~\eqref{eq.a0inv} 
and \eqref{eq.ggpeinv}, it turns out that with real $c_1$ and $c_2$ we are able to give a good 
description of the event distribution, and with  $c_2'$ alone one can reasonably reproduce 
the cross section. We verify that freeing the parameter $c_1'$ barely changes the fit quality. 
Therefore we fix $c_1'=0$ as in the LO case. 

 \begin{figure}[htbp]
   \centering
   \includegraphics[width=0.7\textwidth,angle=-0]{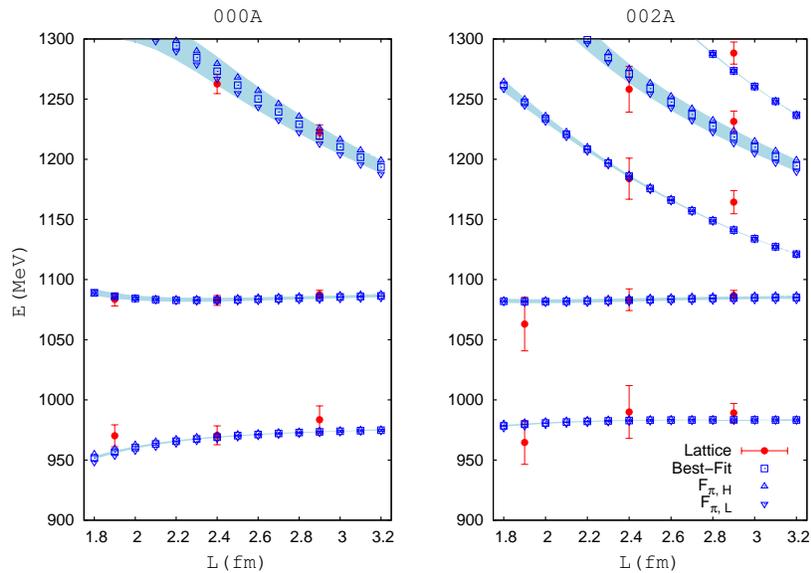} 
  \caption{Fit results of the finite-volume energy levels for the ensembles $000A$ and $002A$ 
           at next-to-leading order. For notations, see
           Fig.~\ref{fig.lolev000}. }
   \label{fig.nlolev000}
\end{figure} 
 
 \begin{figure}[htbp]
   \centering
   \includegraphics[width=0.8\textwidth,angle=-0]{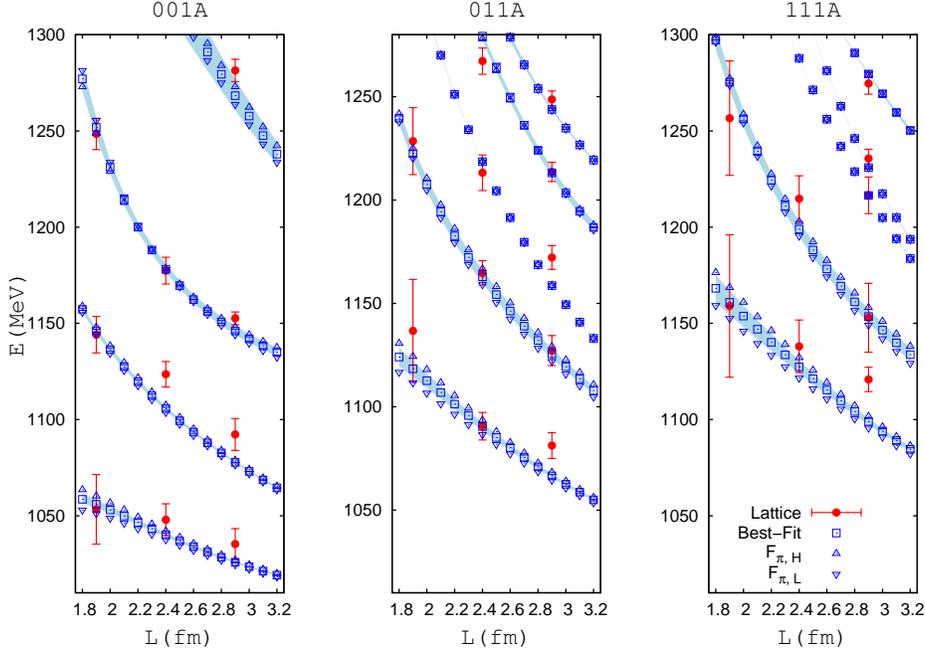} 
  \caption{Fit results of the finite-volume energy levels for the ensembles  $001A$, $011A$ and $111A$ 
           at next-to-leading order. For notations, see Fig.~\ref{fig.lolev000}.} 
   \label{fig.nlolev001}
\end{figure}

The best NLO fit gives $\chi^2/d.o.f = 105.4/(68-5) \simeq 1.67$, among which 72.7 is contributed 
by the lattice energy levels. The values of the two subtraction constants are 
\begin{eqnarray} 
  a_{\pi\eta}= 0.56\pm0.90\pm0.05\,, \qquad a_{K\bar{K}}= -1.62\pm0.33\pm0.02\,. 
\end{eqnarray} 
Note that within errors the present determination of the subtraction constant $a_{\pi\eta}$  
agrees with the value in Ref.~\cite{Guo:2011pa}, which gives $a_{\pi\eta}\simeq 2\pm 3$. 
The subtraction constant $a_{K\bar K}$ in $IJ=10$ scattering in Ref.~\cite{Guo:2011pa} was simply taken 
from the $IJ=00$ $\pi\pi$ scattering:  $a_{K\bar{K}}\simeq -1.15\pm 0.1$, by invoking $SU(3)$ symmetry. 
The values for the parameters related to the $\pi\eta$ production are:  
$c_1=0.48 \pm 0.16\pm0.01$~MeV${^{-1}}$, $c_2=-0.34 \pm 0.10\pm0.00$~MeV${^{-1}}$ and 
$c_2'=2.22\pm0.50\pm0.01$. The first error bar of each parameter is statistical and the second one 
is given by the uncertainties of the unphysical masses in Eq.~\eqref{eq.masslat}. The error bars of 
the parameters are calculated in the same way as explained in the LO case. The values of the 
parameters involved in the $\pi\eta$ production reactions turn out to be rather similar to 
their LO values, indicating that the $D$ functions in Eq.~\eqref{eq.a0inv} for both fits corresponding to the 
$a_0(980)$ are not very different. It is worth pointing out that we have tried to fit only 
the lattice energy levels with the NLO unitarized amplitudes. Though the $\chi^2$ from the 
lattice energy levels in this case decreases around 10 compared with the corresponding value 
from the global fit, the resulting NLO amplitudes turn out to be rather different from the LO 
ones and they give unsatisfactory descriptions of the experimental $\pi\eta$ event distribution 
data within the simple formalism in Eq.~\eqref{eq.a0inv}, even with the complex $c_1$ parameter. 
Moreover, the well-established $a_0(1450)$ resonance, which is explicitly introduced in the NLO 
amplitude, is also strongly distorted, i.e. far away from its PDG value~\cite{Olive:2016xmw}, which 
hints that the NLO fit to only lattice energy levels does not seem to correspond to the real 
physical solution. Therefore we refrain from discussing this fit further and focus on the 
global fit by simultaneously including the lattice and the experimental data. In this respect, 
we argue that the lattice energy levels  with smaller pion masses can be quite useful to 
further constrain the NLO amplitudes. This is because unlike the description of the complicated 
$\pi\eta$ production, no additional theoretical uncertainties will be introduced to study the 
lattice energy levels, as they are purely determined by the $\pi\eta$ scattering.

The NLO fit results for the lattice energy levels are given in Figs.~\ref{fig.nlolev000} and 
\ref{fig.nlolev001}. The  reproduction of the $\pi\eta$ event distribution and 
$\gamma\gamma\to\pi\eta$  cross section is shown in Figs.~\ref{fig.eventdis} and 
\ref{fig.crosssection}, respectively, together 
with the LO results. The meaning of the symbols used in the LO figures is kept for the 
NLO ones. Though the overall reproduction of the lattice energy levels and experimental data is 
quite similar for the LO and NLO fits, the latter gives a slightly better description of the 
47 lattice energy levels with $\chi^2=72.7$ than the former case with $\chi^2=90.2$. The 
LO fit yields better results for the $\pi\eta$ event distribution and the $\gamma\gamma\to\pi\eta$ 
cross section. In other words, the NLO unitarized amplitude seems to work better for the $\pi\eta$ 
dynamics at large unphysical masses, while the LO amplitude seems more efficient to reproduce 
the experimental peaks around the $a_0(980)$ region. Nevertheless, this statement should be 
taken with a grain of salt, because many parameters in the NLO amplitudes are determined from 
other processes in Ref.~\cite{Guo:2011pa} and they can not be solely fixed by the current 
available data from $\pi\eta$ scattering. A more refined theoretical model to describe 
complicated $\pi\eta$ production mechanisms to fit experimental data might also be useful to 
further discern the two amplitudes.

\section{Phase shifts, inelasticities and poles at unphysical masses}\label{sec.latdiscuss}

\begin{figure}[htbp]
   \centering
   \includegraphics[width=0.8\textwidth,angle=-0]{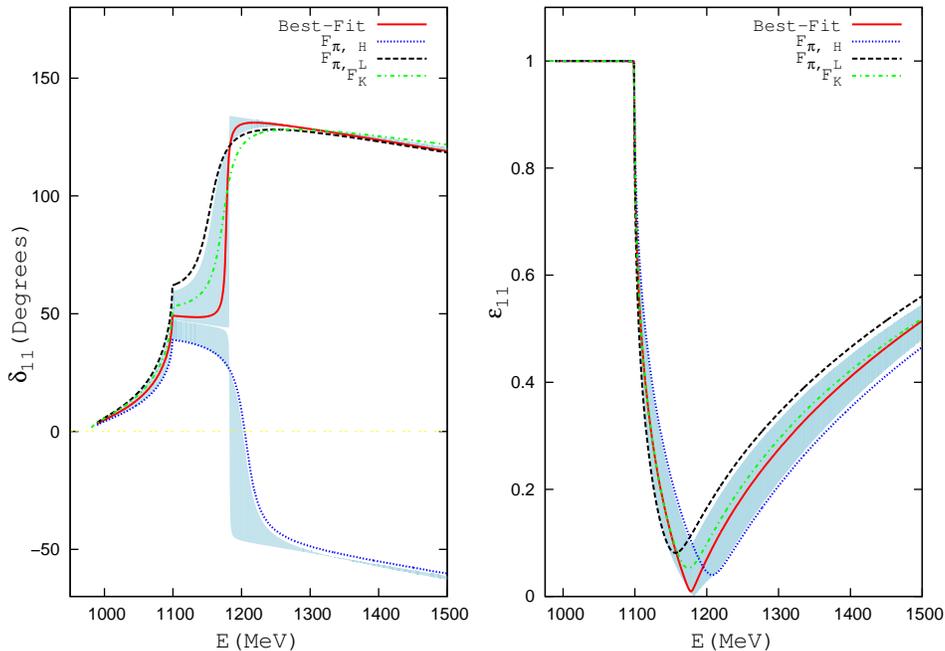} 
  \caption{Leading-order results for phase shifts and inelasticities for $\pi\eta \to \pi\eta$ 
           scattering with the unphysical masses for the $\pi,K,\eta$ and $\eta'$  given in 
           Eq.~\eqref{eq.masslat}. The best-fit results are plotted as red solid lines and the 
           shaded areas represent the 1 $\sigma$ uncertainties. The blue dotted (denoted by 
           $F_{\pi,{\rm H}}$) and the black dashed (denoted by $F_{\pi,{\rm L}}$) lines show the results 
           by taking the upper and lower limits of $F_\pi$ in Eq.~\eqref{eq.fpilaterr}, respectively. 
           The green dashed-dotted lines correspond to the results by distinguishing between 
           $F_\pi$ and $F_K$ in the scattering amplitudes. See the text for more details. }
   \label{fig.lophase}
\end{figure}

\begin{figure}[htbp]
   \centering
   \includegraphics[width=0.8\textwidth,angle=-0]{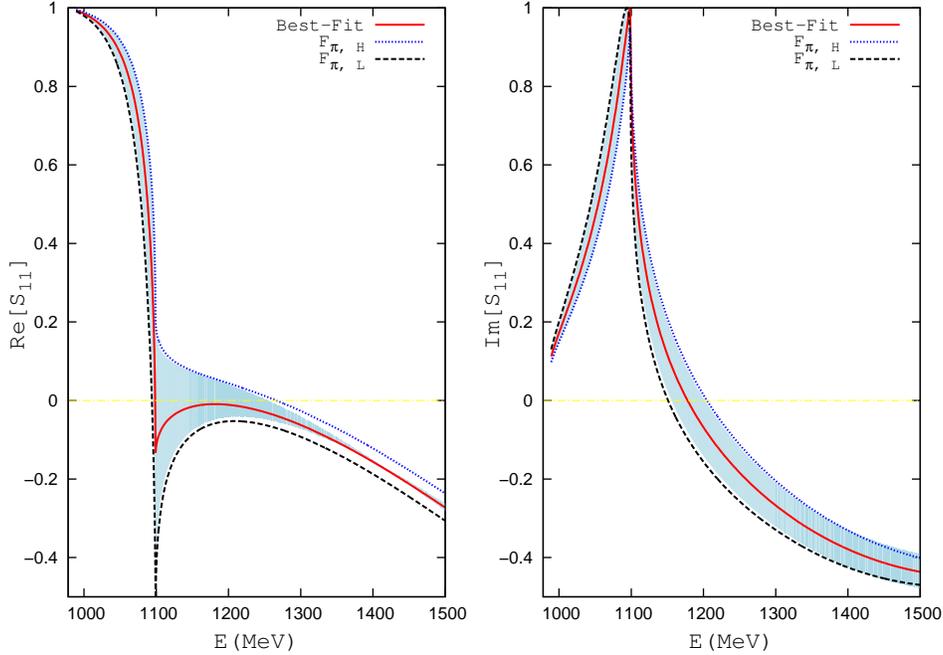} 
  \caption{ The $S$ matrix defined in Eq.~\eqref{eq.defsmat} 
            for $\pi\eta \to \pi\eta$ scattering at leading order with heavy unphysical meson masses in Eq.~\eqref{eq.masslat}. 
            The left panel is for the real part, and the right panel is for the imaginary part. 
            For notations, see Fig.~\ref{fig.lophase}.  }
   \label{fig.losmat}
\end{figure}

\begin{figure}[htbp]
   \centering
   \includegraphics[width=0.8\textwidth,angle=-0]{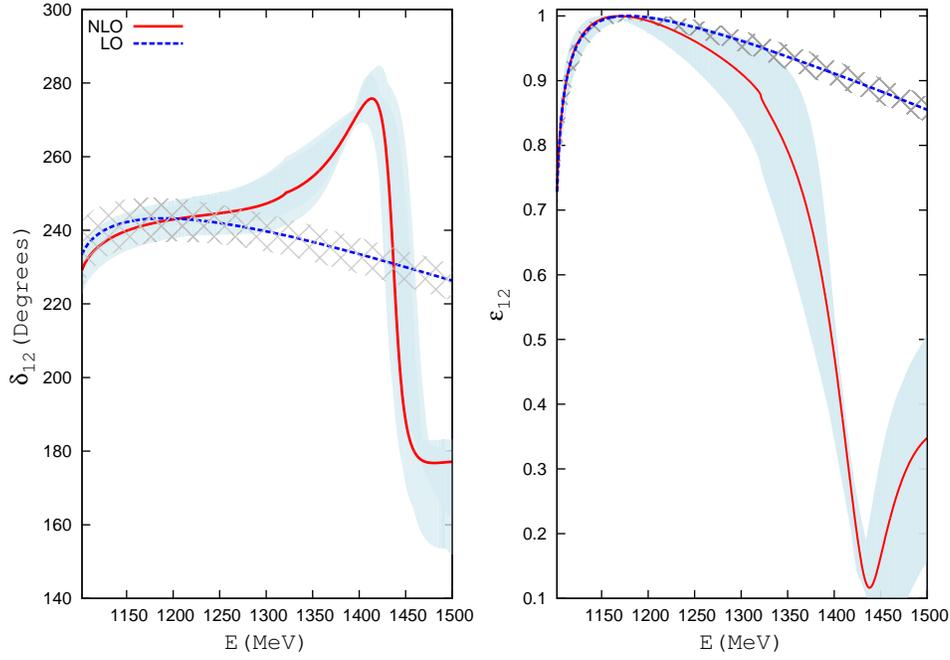} 
  \caption{Phase shifts and inelasticities for  $\pi\eta \to K\bar{K}$ scattering with the 
           heavy unphysical masses for $\pi,K,\eta$ and $\eta'$ in Eq.~\eqref{eq.masslat}. 
           The left panel is for the phase shifts and the right one is for inelasticities. 
           Both the LO (blue dashed lines) and NLO (solid red lines)  results are given. 
           The shaded areas represent the 1-$\sigma$ uncertainties.    }
   \label{fig.latphase12}
\end{figure}

After fixing the parameters in the unitarized chiral amplitudes, we calculate the phase shifts, 
inelasticities, resonance poles and their residues with the unphysical masses used 
in Ref.~\cite{Dudek:2016cru}.  
In Fig.~\ref{fig.lophase}, we give our LO predictions for the $\pi\eta$ scattering phase shifts and 
inelasticities. We observe two different kinds of solutions for the phase shifts within 1 $\sigma$ 
uncertainty. Both of them show a clear kink structure at the $K\bar{K}$ threshold. 
On the one hand, for the first set of solutions we see that the phase shifts show a steep increase 
around 1.2~GeV and are always positive. We explicitly verify that this solution corresponds to the 
situation when $a_{\pi\eta} < -1.41$. Our best fit and the result with a lower limit of $F_\pi$ in 
Eq.~\eqref{eq.fpilaterr}, shown with red solid and black dashed lines in Fig.~\ref{fig.lophase}, 
respectively, belong to this kind of solution. 
On the other hand, in the second case when $a_{\pi\eta} > -1.41$ the phase shifts exhibit mild and 
continuous changes with increasing energies and become negative in the energy region above 1.2~GeV. 
The result obtained with the upper limit of $F_\pi$ in Eq.~\eqref{eq.fpilaterr} is similar 
to the second case. It is also interesting to note that the phase shifts obtained in the lattice 
analyses in Ref.~\cite{Dudek:2016cru} are similar to our second type of solution, i.e., to the lower 
branch shown in the left panel of  Fig.~\ref{fig.lophase}. 
In the right panel, we give the inelasticity of the $\pi\eta$ scattering. 
Below the $K\bar{K}$ threshold, the inelasticity is equal to 1, as it should be. 
At the $K\bar{K}$ threshold the inelasticity suddenly decreases to almost zero and gradually increases 
when the energy becomes larger. As shown in Fig.~\ref{fig.lophase}, the inelasticities show a qualitatively 
similar behavior within 1-$\sigma$ uncertainty and with different extrapolation forms of $F_\pi$. 
On physical grounds, both types of solutions for the phase shifts are indeed very similar, since above 
1.2~GeV both results for the $\pi\eta$ phase shifts only differ by 180 degrees. Although in the energy 
region between the $K\bar{K}$ threshold and 1.2~GeV the phase shifts show large uncertainties, the 
inelasticity in this region is almost zero. In order to clearly demonstrate the similarity of the 
underlying dynamics between these two different branches of phase shifts in Fig.~\ref{fig.lophase}, 
we give the $S$ matrix for the $\pi\eta\to \pi\eta$ scattering in Fig.~\ref{fig.losmat}. 
This also indicates that the $\pi\eta\to K \bar{K}$ scattering plays a more important role in 
this specific energy range. In Fig.~\ref{fig.latphase12}, we show the LO phase shifts 
(left panel) and inelasticities (right panel) for the $\pi\eta\to K \bar{K}$ scattering with 
blue dashed lines. Note that, as expected, this transition amplitude just varies slightly 
within the 1 $\sigma$ region. 
It is worth pointing out that around the $K\bar{K}$ threshold the $\pi\eta$ phase shifts 
are also found to be quite sensitive to the small variation of parameters in 
Ref.~\cite{Dudek:2016cru}, but the scattering amplitudes are stable.

\begin{figure}[htbp]
   \centering
   \includegraphics[width=0.8\textwidth,angle=-0]{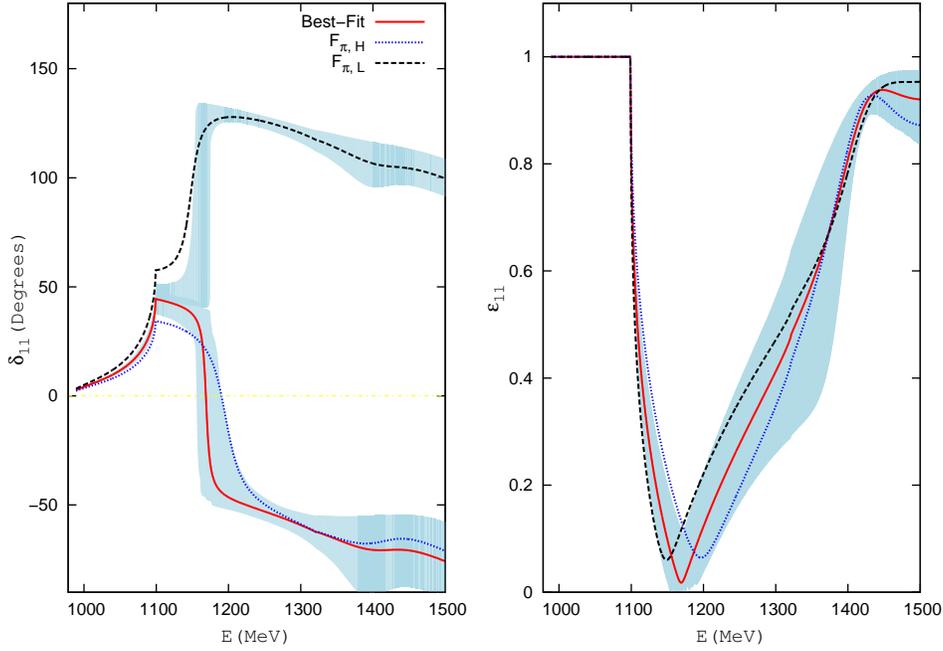} 
  \caption{NLO results for phase shifts and inelasticities for $\pi\eta \to \pi\eta$ scattering 
           with the unphysical masses for $\pi,K,\eta$ and $\eta'$ in Eq.~\eqref{eq.masslat}. 
           For notations, see Fig.~\ref{fig.lophase}. }
   \label{fig.nlophase}
\end{figure}

\begin{figure}[htbp]
   \centering
   \includegraphics[width=0.8\textwidth,angle=-0]{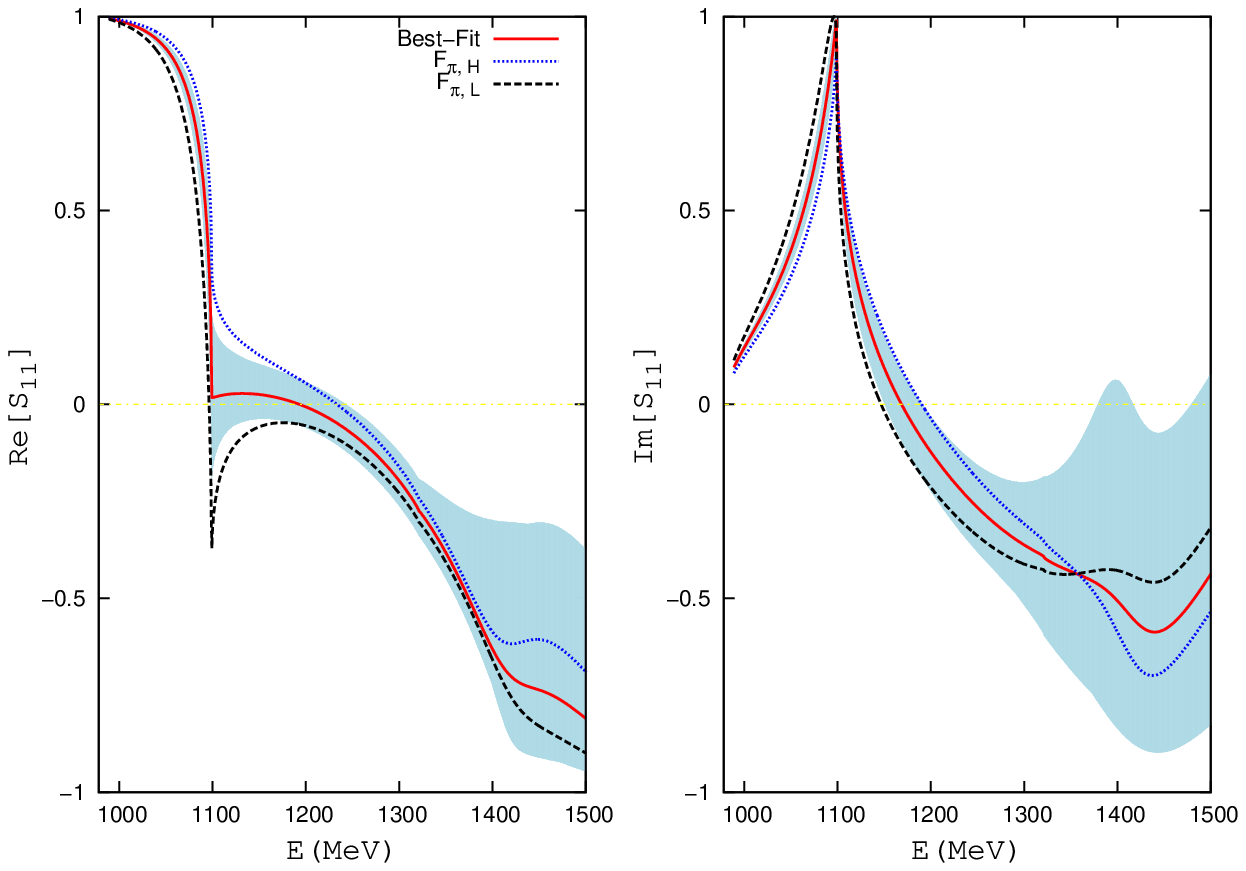} 
  \caption{ The $S$ matrix defined in Eq.~\eqref{eq.defsmat}  for $\pi\eta \to \pi\eta$ scattering at next-to-leading order with heavy unphysical meson masses in Eq.~\eqref{eq.masslat}. The left (right) panel is for the real (imaginary) part. For notations, see Fig.~\ref{fig.lophase}.   }
   \label{fig.nlosmat}
\end{figure}

Our NLO predictions for the $\pi\eta$ scattering phase shifts and inelasticities 
calculated at the unphysical masses of Eq.~\eqref{eq.masslat} are shown in Fig.~\ref{fig.nlophase}. 
Similarly to  the LO case, two different kinds of solutions for the phase shifts within 1-$\sigma$ 
uncertainty are found. 
The first set of fits for the phase shifts shows a steep increase around 1170~MeV and are always positive.  
The result with the lower limit of $F_\pi$ in Eq.~\eqref{eq.fpilaterr}, shown by the black dashed line in 
Fig.~\ref{fig.nlophase} belongs to this kind of solution. Most of the parameter configurations of the NLO 
fits lead to the second type of solution: 
The phase shifts exhibit mild and continuous changes with increasing energies and become negative in the 
energy region above  1170~MeV or so. The result obtained with the upper limit of $F_\pi$ in 
Eq.~\eqref{eq.fpilaterr} belongs to the second type of solution. The inelasticities of the 
$\pi\eta$ scattering are given in the right panel of Fig.~\ref{fig.nlophase}. 
The NLO inelasticities show a different behavior compared to the LO ones above the 1.3~GeV region. 
There is a rapid increase in the NLO case around 1.4~GeV. 
The reason behind this behavior is that in the NLO amplitude the $a_0(1450)$ resonance 
is explicitly included,  while only the lowest order 
contact meson-meson interactions are incorporated  at LO 
and the $a_0(1450)$ can not be generated in this case. 
Similar to the LO situation, the phase shifts above 1.2~GeV only differ by 180 degrees, but show 
large uncertainties in the energy range between the $K\bar{K}$ threshold and 1.2~GeV. However, these 
large uncertainties do not matter since the  inelasticities are very small in the same region. 
This statement can be clearly seen in Fig.~\ref{fig.nlosmat}, where the real and imaginary parts 
of the $S$ matrix for $\pi\eta\to\pi\eta$ scattering at NLO are displayed.  
Then the physics is dominated by the $\pi\eta\to K\bar{K}$ scattering in this region. The NLO 
phase shifts and inelasticities for $\pi\eta \to K\bar{K}$ scattering, together with the LO results, 
are given in Fig.~\ref{fig.latphase12}. One can clearly see that the LO and NLO phase shifts and 
inelasticities for $\pi\eta \to K\bar{K}$ scattering are quite similar in the range from the 
$K\bar{K}$ threshold up to around 1.3~GeV, somewhat before the effects of the $a_0(1450)$ 
resonance become dominant.

A unique way to characterize a resonance is to look for the corresponding poles in the complex energy plane. 
This is also the only model-independent method. 
In our framework, one can extrapolate to the complex energy plane by modifying the 
infinite-volume $G$ function in Eq.~\eqref{eq.gfuncdr}. Three two-body thresholds, 
i.e. $\pi\eta, K\bar{K}$ and $\pi\eta'$, 
introduce $2^3$ RS's in the complex plane. 
The $G$ function for  each channel has two RS's and the expression in Eq.~\eqref{eq.gfuncdr} 
corresponds to the first RS. Its expression on the second RS takes the form 
\begin{eqnarray}\label{eq.defg2ndrs}
 G(s)^{\rm DR}_{II}(s) = G(s)^{\rm DR} + i \frac{\sigma(s)}{8\pi s}\,,
\end{eqnarray}
with $G(s)^{\rm DR}$ and $\sigma(s)$ defined in Eqs.~\eqref{eq.gfuncdr} and \eqref{eq.defsigkin}, 
respectively.\footnote{In the 
complex $s$ plane, $\sigma(s)$ has to be calculated such that $\text{Im}\,\sigma(s)>0$.} 
Changing from the first RS to the second one implies reversing the sign of the imaginary part of the 
$G$ function along the real $s$ axis above threshold. 

We denote the physical/first RS by $(+,+,+)$, where the plus sign in each entry indicates that 
the $G$ function is evaluated in the physical RS at $\pi\eta, K\bar{K}$ and $\pi\eta'$ thresholds, in order. 
The second, third, fourth and fifth RS are labeled as $(-,+,+)$, $(-,-,+)$, $(+,-,+)$ and 
$(-,-,-)$, respectively, with the minus sign indicating that the $G$ function for this channel is 
evaluated in its second RS, cf.~Eq.\eqref{eq.defg2ndrs}. The same convention has also been used in 
Refs.~\cite{Guo:2011pa,Dudek:2016cru}, making the comparison between different approaches 
straightforward. In addition to the pole positions, we also calculate the residues for the 
three different channels, which characterize the couplings of the poles to the different channels.

Both for the LO and NLO cases, we find one relevant pole near the $K\bar{K}$ threshold, which is 
located either on the second or fourth sheet within 1 $\sigma$ uncertainty. In fact, we find that 
the poles in the second sheet correspond to the parameter configurations that lead to the upper branch 
of phase shifts in Figs.~\ref{fig.lophase} and \ref{fig.nlophase}, whereas the poles in the fourth 
sheet correspond to the parameters that give the lower branch of the phase shifts in 
Figs.~\ref{fig.lophase} and \ref{fig.nlophase}. The relations between the transition of 
pole locations and the different behaviors of phase shifts are also noticed in Ref.~\cite{Dudek:2016cru}. 
The explicit values of the pole positions, together with their residues, are given in 
Table~\ref{tab.polelat}. Notice that the central values of parameters of the LO and NLO fits 
lead to the $a_0(980)$ poles on the second and fourth sheets, respectively. At NLO, the poles around 
the $K\bar{K}$ threshold are quite similar to those at LO,
though the masses of both the second- and fourth-sheet poles in the NLO amplitude are about 
20~MeV below the LO ones.  We also note that the couplings to the $\pi\eta'$ channel for the  
poles around the $K\bar{K}$ threshold in both LO and NLO cases are small, implying a marginal 
role of this channel 
when determining the $a_0(980)$ state. Let us note  that no pole for the $a_0(980)$ in 
Table~\ref{tab.polelat} lies in the unphysical RS that matches the physical RS above the 
$K\bar{K}$ threshold, so only the  low-energy tail of the pole 
in the second RS is directly realized on the real energy axis below this threshold. 
One obvious difference between the LO and NLO amplitudes is that the latter contains a resonance pole 
located at around $1420$~MeV in the fifth RS, corresponding to the $a_0(1450)$, which is absent in 
the LO case. The $\pi\eta'$ channel is found to be important for the heavy $a_0(1450)$ resonance, 
since the coupling to the $\pi\eta'$ channel is even larger than to the coupling to $\pi\eta$ one for this resonance, 
as shown in Table~\ref{tab.polelat}.

Next we make a brief comparison with the pole content in Ref.~\cite{Dudek:2016cru}. Around the 
$K\bar{K}$ threshold region, one fourth-sheet pole is found, with mass $1177 \pm 27$~MeV 
and width $ 49 \pm 33$~MeV.  The error bars were obtained by averaging many different types 
of parametrizations in Ref.~\cite{Dudek:2016cru}. 
In our case the $a_0(980)$ pole can be either on the second or fourth sheet within 1 $\sigma$ 
uncertainty. The pole content resulting from the global fits, by including both the lattice and 
experimental data, is summarized in Table~\ref{tab.polelat}. In order to make a more clear 
comparison with Ref.~\cite{Dudek:2016cru}, we also give the pole contents from the LO fit by only 
including the 47 lattice energy levels. Again, the pole can be located either on the second or the 
fourth sheet within 1 $\sigma$ uncertainty. The mass and width on the second sheet are  
$1170_{-26}^{+12}$~MeV and $16_{-16}^{+34}$~MeV, respectively. The central value of the mass and 
width of the second sheet pole simply corresponds to taking the central value of the subtraction 
constant from this fit. For the mass and width of the fourth-sheet pole, we take the 
median numbers as their central values. Then the mass and width of the pole on the fourth sheet 
are $1192_{-10}^{+11}$ and $12_{-12}^{+12}$~MeV. Similar rules are also applied to other numbers 
in Tables~\ref{tab.polelat} and \ref{tab.polephy}. As pointed out in Ref.~\cite{Dudek:2016cru} 
the lower half-plane of the second sheet is continuously connected to the upper half of the 
fourth sheet, which indicates that the nearby pole in the second  or fourth sheet in fact 
represents quite similar physics.  In Fig.~\ref{fig.losmat} we further confirm this conclusion: 
The $S$ matrix exhibits continuous changes within uncertainties, though different sheets of poles are found.

Within uncertainties either a fourth-sheet virtual pole ranging from 971 to 978~MeV near the $\pi\eta$ 
threshold or a third-sheet virtual pole ranging from 975 to 978~MeV  are found for the LO case, 
which confirms the result in Ref.~\cite{Dudek:2016cru}, giving $964 \pm 62$~MeV. However, at NLO 
we find only a prominent bump around $976$~MeV in the fourth sheet, instead of a pole. The 
fourth- or third-sheet virtual pole does not produce a prominent structure for the $\pi\eta$ scattering 
amplitude on the physical axis. Other poles that are far away from the $K\bar{K}$ threshold in the 
third sheet are also found in Ref.~\cite{Dudek:2016cru} and in our case. Since these poles are 
so far away from the energy region we are focusing on, we do not discuss them any further.

\begin{table}[htbp]
 \centering
{\small 
\begin{tabular}{ c c | c c c c c}
\hline\hline
Resonance  & RS & Mass   & Width/2   & $|{\rm Residue}|_{\pi\eta}^{1/2}$ & Ratios & \\
   & & (MeV) &  (MeV) & (GeV) &   & 
\\ \hline
LO  &  & &   &   &  & \\
$a_0(980)$ & II & $1178_{-20}^{+4}$ & $3_{-3}^{+13}$ & $5.6_{-1.6}^{+0.1}$
 & $1.23_{-0.01}^{+0.04}$ ({\footnotesize $K\bar{K}/\pi\eta$}) & $0.18_{-0.01}^{+0.02}$ ({\footnotesize $\pi\eta'/\pi\eta$})\\ 
$a_0(980)$ & IV & $1189_{-6}^{+15}$ & $4_{-4}^{+9}$ & $5.8_{-1.5}^{+0.3}$
 & $1.21_{-0.03}^{+0.01}$ ({\footnotesize $K\bar{K}/\pi\eta$}) & $0.16_{-0.02}^{+0.01}$ ({\footnotesize $\pi\eta'/\pi\eta$})\\ 
 \hline
NLO  &  & &   &   &  & \\
$a_0(980)$  & II & $1160_{-10}^{+14}$ & $2_{-2}^{+5}$ & $3.6_{-0.5}^{+0.9}$ 
 & $1.29_{-0.03}^{+0.04}$ ({\footnotesize $K\bar{K}/\pi\eta$}) & $0.19_{-0.01}^{+0.00}$ ({\footnotesize $\pi\eta'/\pi\eta$})\\ 
$a_0(980)$  & IV & $1169_{-13}^{+26}$ & $4_{-4}^{+16}$ & $4.4_{-1.0}^{+1.4}$
 & $1.25_{-0.05}^{+0.04}$ ({\footnotesize $K\bar{K}/\pi\eta$}) & $0.19_{-0.01}^{+0.01}$ ({\footnotesize $\pi\eta'/\pi\eta$})\\
$a_0(1450)$  & V & $1418_{-15}^{+13}$ & $54_{-18}^{+70}$ & $1.0_{-0.1}^{+0.8}$
 & $2.9_{-0.9}^{+1.2}$ ({\footnotesize $K\bar{K}/\pi\eta$}) & $1.8_{-0.6}^{+0.3}$ ({\footnotesize $\pi\eta'/\pi\eta$})\\
 \hline\hline
\end{tabular}
\caption{ Pole positions and the corresponding residues when the masses of pNGBs are taken at their lattice 
          values in Eq.~\eqref{eq.masslat}. The thresholds of $\pi\eta$, $K\bar K$ and $\pi\eta'$ are $978.5$, 
          $1099$, and $1321.1$~MeV, respectively.  We point out that there is only one pole around the $K\bar{K}$ threshold for each parameter configuration. Nevertheless, within 1 $\sigma$ uncertainty different parameter configurations can either give a pole on the second sheet or the fourth sheet. See the text for details. }\label{tab.polelat}}
\end{table}

\section{Phase shifts, inelasticities and poles at the physical masses}\label{sec.phydiscuss}

Since $U(3)$ $\chi$PT is based on the chiral symmetry of QCD, it provides a useful framework 
to perform the chiral extrapolation from unphysically large 
pion masses to its physical value. Therefore, in this section, we give the predictions for the phase shifts, 
inelasticities, pole positions, and the residues for the $\pi\eta$ scattering by taking the 
physical masses for the $\pi, K, \eta$ and $\eta'$ mesons. 

\begin{figure}[htbp]
   \centering
   \includegraphics[width=0.8\textwidth,angle=-0]{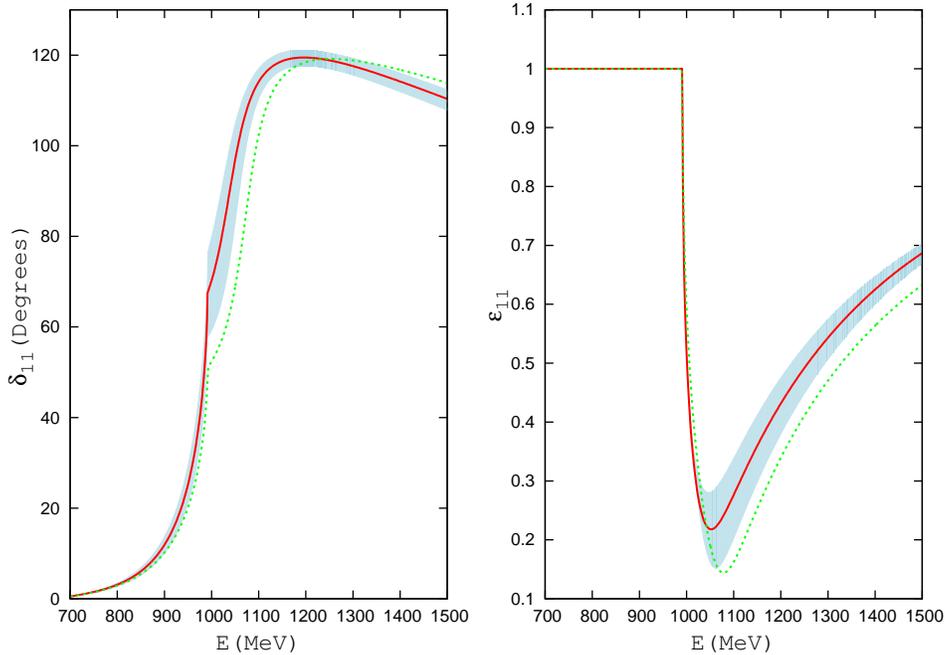} 
  \caption{Phase shifts and inelasticities from $\pi\eta \to \pi\eta$ scattering with physical masses 
           at leading order. The red solid lines correspond to the results from the best fit using a common pion decay constant in all the amplitudes. The shaded areas represent the statistical 1-$\sigma$ uncertainties. The green dotted lines denote the results by distinguishing between $F_\pi$ and $F_K$ in the scattering amplitudes. See the text for details. }
   \label{fig.lophasephy}
\end{figure}

\begin{figure}[htbp]
   \centering
   \includegraphics[width=0.8\textwidth,angle=-0]{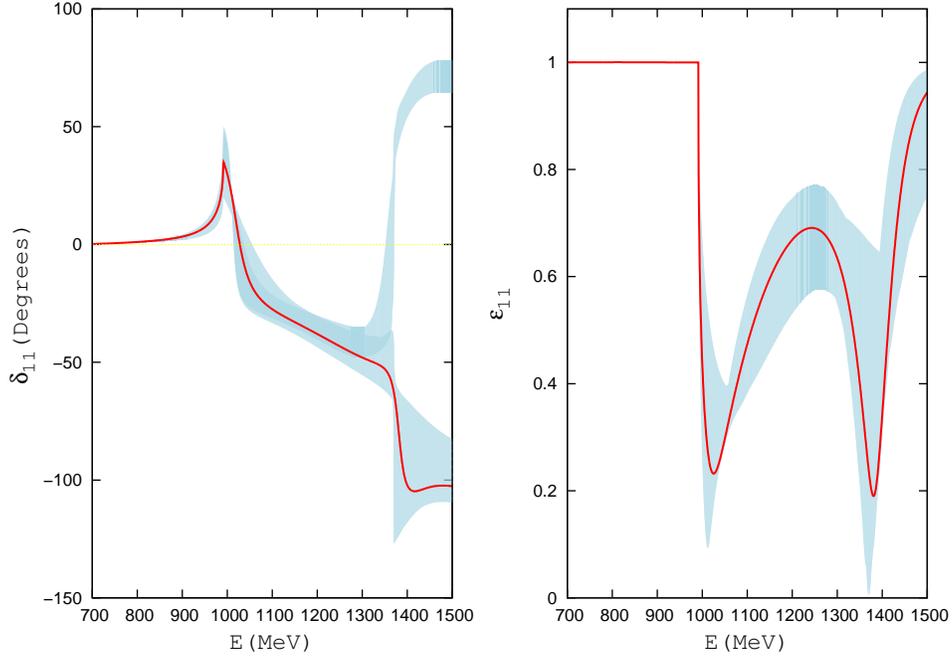} 
  \caption{Phase shifts and inelasticities from $\pi\eta \to \pi\eta$ scattering with physical masses 
           at next-to-leading order.  The shaded areas represent the statistical 1-$\sigma$ uncertainties.}
   \label{fig.nlophasephy}
\end{figure}

\begin{figure}[htbp]
   \centering
   \includegraphics[width=0.8\textwidth,angle=-0]{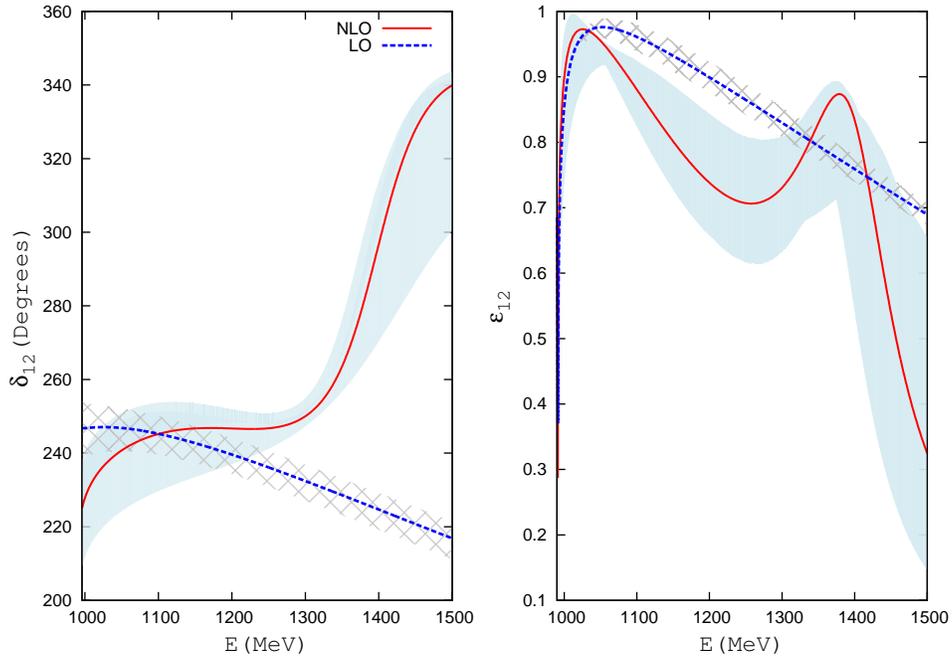} 
  \caption{Phase shifts and inelasticities for the $\pi\eta \to K\bar{K}$ scattering with the physical 
           masses for $\pi,K,\eta$ and $\eta'$. For notations, see 
           Fig.~\ref{fig.latphase12}.  }
   \label{fig.phyphase12}
\end{figure}

As in the previous section, we present the results for the LO and NLO study separately. 
The LO predictions for the $\pi\eta$ phase shifts and inelasticities are shown in the left and 
right panels of  Fig.~\ref{fig.lophasephy}, respectively. 
The corresponding predictions at NLO  are given in Fig.~\ref{fig.nlophasephy}.  
We observe very different results by comparing the two figures.  
Unlike the unphysical mass case, only one solution is found for LO. Although one solution is 
found for NLO around the $K\bar{K}$ threshold, two branches of phase shifts within 1 $\sigma$ 
uncertainty appear at NLO in the energy region above around 1.4~GeV. 
To be more specific, we always observe a steep increase around the $K\bar{K}$ threshold for the 
LO $\pi\eta$ phase shifts, 
while the NLO phases continuously decrease above the $K\bar{K}$ threshold until the appearance of 
the $a_0(1450)$.  In the energy region around 1.4~GeV, we find large uncertainties for the NLO phase 
shifts. However, the inelasticities turn out to be quite small in the same energy range. Then the 
situation here is similar to the discussions about large unphysical meson masses around the 1.2~GeV 
region in Figs.~\ref{fig.lophase} and \ref{fig.nlophase}.  
Due to the inclusion of the $a_0(1450)$, more complicated structures for the inelasticities 
appear at NLO than at LO. We also give the phase shifts and inelasticities from the $\pi\eta \to 
K\bar{K}$ scattering both at LO and NLO evaluated with physical masses in Fig.~\ref{fig.phyphase12}.  
We point out that the uncertainties given in  Figs.~\ref{fig.lophasephy}, \ref{fig.nlophasephy} 
and \ref{fig.phyphase12} should be taken with caution since only the statistical errors are 
included here; the systematic errors caused by the theoretical uncertainties and the 
chiral extrapolations are not considered. 
One possible theoretical uncertainty is given by using different pNGB decay constants in the 
scattering amplitudes. We make an exploratory study about this effect. The phase shifts and inelasticities 
with the replacement of $F_\pi$ by $F_K$ in the amplitudes involving kaons are shown as green 
dashed-dotted and dotted lines in Figs.~\ref{fig.lophase} and \ref{fig.lophasephy}, respectively. 
Quantitatively results similar  to those from using a common $F_\pi$ in all the amplitudes are observed. 

The relevant pole positions for the $a_0(980)$ and $a_0(1450)$ in the complex energy plane and the 
corresponding residues are given in Table~\ref{tab.polephy}.
Only one second RS pole for the  $a_0(980)$ is found in the LO case, while one pole located on the 
fourth RS is found in the NLO amplitude. The $a_0(980)$ poles in both cases are clearly above the 
$K\bar{K}$ threshold, and they are found to be barely coupled to the $\pi\eta'$ channel. Neither of them 
lies in the RS that matches with the physical sheet in the energy interval between the $K\bar{K}$ 
and $\pi\eta'$ thresholds. But, while for the LO case the pole in the second RS is directly 
accessible from the energy 
axis below this threshold, this is not the case for the hidden fourth RS pole in the NLO case. 
The most relevant pole for $a_0(1450)$ is located on the fifth RS since it lies above the 
$\pi\eta'$ threshold. The coupling strength of the $a_0(1450)$ to the $\pi\eta'$ channel is found to 
be similar to the $\pi\eta$ one, and therefore should be included when discussing this excited $a_0$ state. 
We mention that other redundant poles are also found in our unitarized amplitudes, such as a third-sheet 
pole with mass around $750$~MeV and width around $100$~MeV. However, the redundant poles, which are 
usually located in the position that is not directly connected to the physical RS, do not show any 
visible effects on the physical axis. Therefore we refrain from discussing them any further.

Our current predictions for the $\pi\eta$ phase shifts are different from the recent study 
in Ref.~\cite{Albaladejo:2015aca}.  The reason for this is not difficult to understand. 
In that work, two specific pole positions for the $a_0(980)$ in the second and third RS's, 
and one pole for the $a_0(1450)$ in the third RS are taken as external input to determine the phase shifts. 
In other words, the phase shifts given in Ref.~\cite{Albaladejo:2015aca} are (at least partially) determined 
{\it a priori} by the imposed pole positions of the $a_0(980)$ and $a_0(1450)$. 
This is clearly different from our method, since the pole positions in Table~\ref{tab.polephy} are not 
imposed beforehand. Instead, our pole content is determined once the phenomenological and lattice 
inputs are successfully reproduced. Indeed,  we do not find any third RS pole for the $a_0(980)$ 
in our study, while in Ref.~\cite{Albaladejo:2015aca}  this kind of pole is imposed to find the 
phase shifts. 
In our NLO study, we do not find any second RS pole and only one pole in the fourth RS is found. 
With different pole content embedded in the chiral amplitudes, it is not surprising to observe 
different solutions for the phase shifts. The phase shifts and inelasticities obtained here can 
provide important inputs for the dispersive study of processes involving $\pi\eta$~\cite{Escribano:2016ntp,Albaladejo:2016mad}.

\begin{table}[htbp]
 \centering
 {\small 
\begin{tabular}{ c c | c c c c c}
\hline\hline
Resonance  & RS & Mass   & Width/2   & $|{\rm Residue}|_{\pi\eta}^{1/2}$ & Ratios & \\
   & & (MeV) &  (MeV) & (GeV) &   & 
\\ \hline
LO  &  & &   &   &  & \\
$a_0(980)$ & II & $1037_{-14}^{+17}$ & $44_{-9}^{+6}$ & $3.8_{-0.2}^{+0.3}$
 & $1.43_{-0.03}^{+0.03}$ ({\footnotesize $K\bar{K}/\pi\eta$}) & $0.05_{-0.01}^{+0.01}$ ({\footnotesize $\pi\eta'/\pi\eta$})\\ 
 \hline
 NLO  &  & &   &   &  & \\
$a_0(980)$ & IV & $1019_{-8}^{+22}$ & $24_{-17}^{+57}$ & $2.8_{-0.6}^{+1.4}$
 & $1.8_{-0.3}^{+0.1}$ ({\footnotesize $K\bar{K}/\pi\eta$}) & $0.01_{-0.01}^{+0.06}$ ({\footnotesize $\pi\eta'/\pi\eta$})\\
$a_0(1450)$ & V & $1397_{-27}^{+40}$ & $62_{-8}^{+79}$ & $1.7_{-0.4}^{+0.3}$
 & $1.4_{-0.6}^{+2.4}$ ({\footnotesize $K\bar{K}/\pi\eta$}) & $0.9_{-0.2}^{+0.8}$ ({\footnotesize $\pi\eta'/\pi\eta$})\\
 \hline\hline
\end{tabular}
\caption{\label{tab.polephy} Pole positions and the corresponding residues when the masses of the pNGBs are fixed 
                             at their physical values given  in Eq.~\eqref{eq.massphy}. The thresholds of the 
                             $\pi\eta$, $K\bar K$, and $\pi\eta'$ channels are $685.2$, $991.2$, and $1095$~MeV, 
                             respectively. }
}
\end{table}

\section{Summary and conclusions}\label{sec.summary}

In this work, we have analyzed very recent lattice finite-volume energy levels in the rest and 
moving frames  for  $\pi\eta$ scattering, together with the experimental data on a $\pi\eta$ 
event distribution  and the $\gamma\gamma\to\pi\eta$ cross section. 
Three coupled channels, 
$\pi\eta$, $K\bar{K}$ and $\pi\eta'$, are considered in our study. 
Both the leading and next-to-leading-order chiral amplitudes are used in the analyses. 
The simultaneous fits to the present lattice QCD
finite-volume levels and the experimental data can not distinguish between 
the leading and next-to-leading-order scenarios, both of which lead to quite similar fit qualities. 

However, 
somewhat different $\pi\eta$ scattering phase shifts are obtained for the leading and 
next-to-leading-order cases,  
when taking the heavy unphysical masses in Eq.~\eqref{eq.masslat}. 
Two branches of solutions for the $\pi\eta$ phase shifts are found within  uncertainties. Nevertheless, 
the two solutions of phase shifts in fact give similar dynamics, when combined with the inelasticities. 
The $\pi\eta \to K\bar{K}$ scattering phase shifts and inelasticities are also provided. 
A pole in either the second or the fourth Riemann sheet is found for the $a_0(980)$ within 
1-$\sigma$ uncertainty, when using the heavy unphysical masses for the $\pi, K, \eta$ and $\eta'$. 
Our determinations for the pole 
of the $a_0(980)$ are compatible with those in Ref.~\cite{Dudek:2016cru} within uncertainties.

The most interesting predictions of this work are given in Sec.~\ref{sec.phydiscuss}. 
The phase shifts and inelasticities of the $\pi\eta\to\pi\eta$ and $\pi\eta\to K\bar{K}$ 
scattering, pole  positions and their residues are calculated by taking the physical masses for 
the $\pi, K, \eta$ and $\eta'$. 
Within the statistical uncertainties, only one set of solutions of the $\pi\eta$ phase shifts is found for 
the leading-order case. Although one set of solutions of the $\pi\eta$ phase shifts is observed at 
next-to-leading order around the $K\bar{K}$ threshold, two branches of solutions are found above 
around 1.4~GeV. For the leading-order scenario, the physical $\pi\eta$ phase shifts clearly show a 
steep increase around the $K\bar{K}$ threshold. However, the phase shifts at next-to-leading order 
decrease continuously above this threshold until the appearance of the $a_0(1450)$ resonance. 
Though at next-to-leading 
order large uncertainties for the $\pi\eta$ phase shifts show up around 1.4~GeV, the inelasticities in 
the same region are quite small. 
The different behaviors of phase shifts are also reflected in the different pole contents. 
One pole slightly above the $K\bar{K}$ threshold is found in the second Riemann sheet for the 
leading-order amplitude (so that its low-energy tail directly influences the amplitudes on the energy 
axis below the $K\bar{K}$ threshold), 
while there is only one hidden fourth-sheet pole in the next-to-leading-order case for the $a_0(980)$.  
Due to the inclusion 
of the $a_0(1450)$ in the next-to-leading-order case, 
which is absent at leading order, the inelasticities from the two orders show different behaviors above 
around 1.1~GeV. The $\pi\eta'$ channel is found to be rather weakly coupled to the $a_0(980)$ at 
both heavy unphysical and physical masses  and hence plays a minor role for the determination of 
the $a_0(980)$ properties. The coupling strength of the $a_0(1450)$ to the $\pi\eta'$ channel is 
nearly as large as  the $\pi\eta$ one.

To summarize,  global fits of similar quality including both experimental and lattice data are obtained, 
using unitarized chiral perturbation theory with two input chiral amplitudes, evaluated at 
leading and next-to-leading order. The leading-order amplitude gives a better description of the 
experimental $\pi\eta$ data evaluated at physical masses, but it gives slightly worse results for the 
lattice energy levels at $m_\pi=391$~MeV. The situation for the next-to-leading-order case is 
just the opposite. More importantly, the two different amplitudes obviously lead to different $\pi\eta$ 
phase shifts for the physical masses. Unlike the $\pi\eta$ experimental data which include the 
complicated $\pi\eta$ production mechanisms, the lattice energy levels are solely determined by 
the $\pi\eta$ scattering information.  It is therefore important to have the finite-volume 
energy levels from the $\pi\eta$ scattering with lighter quark masses in order to discriminate 
between these two different solutions.

\section*{Acknowledgements}

Finite-volume energy levels taken from Ref.~\cite{Dudek:2016cru} were provided by the 
Hadron Spectrum Collaboration -- no endorsement on their part of the analysis presented in 
the current paper should be assumed. 
Z.H.G. would like to thank M.~Albaladejo and J.~J.~Xie for communications, and M.~D\"oring for many 
useful discussions. J.A.O. would like to expresses his gratitude to the HISKP for its kind hospitality during a 
research visit where part of this work was done.  
This work is carried out in the framework of the Sino-German Collaborative 
Research Center ``Symmetries and the Emergence of Structure in QCD'' (CRC~110)
co-funded by the DFG and the NSFC. This work is also supported in part by the NSFC under Grants 
No. 11575052 and 11505038, the Natural Science Foundation of Hebei Province under Contract No.~A2015205205,  
the MINECO (Spain) and ERDF (European Commission) Grant No. FPA2013-40483-P and by the 
Spanish Excellence Network on Hadronic Physics FIS2014-57026-REDT, grant by  VolkswagenStiftung under 
Contract No. 86260. 
The work of U.G.M. was supported in part by The Chinese Academy of Sciences 
(CAS) President's International Fellowship Initiative (PIFI) Grant No. 2017VMA0025.

\end{document}